\documentclass{article}

\usepackage{arxiv}
\usepackage[utf8]{inputenc} 
\usepackage[T1]{fontenc}    
\usepackage{hyperref}       
\usepackage{url}            
\usepackage{booktabs}       
\usepackage{amsfonts}       
\usepackage{nicefrac}       
\usepackage{microtype}      
\usepackage{lipsum}		
\usepackage{graphicx}
\usepackage{natbib}
\usepackage{doi}

\usepackage{amsmath}
\usepackage{rotating}
\usepackage{caption}
\usepackage{bm}
\usepackage{algorithm,algorithmicx,algpseudocode}
\usepackage{booktabs}

\title{Bayesian Semiparametric Mixture Cure (Frailty) Models}

\author{ \href{https://orcid.org/0000-0001-6457-0967}{\includegraphics[scale=0.06]{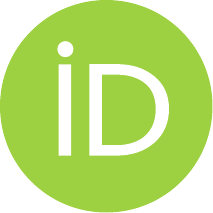}\hspace{1mm}Fatih K\i z\i laslan} \\
	Oslo Centre for Biostatistics and Epidemiology\\
    Department of Biostatistics\\
	University of Oslo\\
	Oslo, Norway\\
	\texttt{fkizilaslan@yahoo.com} \\
	\And
	\href{https://orcid.org/0000-0002-6746-0453}{\includegraphics[scale=0.06]{orcid.pdf}\hspace{1mm}Valeria Vitelli} \\
	Oslo Centre for Biostatistics and Epidemiology\\
    Department of Biostatistics\\
	University of Oslo\\
	Oslo, Norway\\
	\texttt{valeria.vitelli@medisin.uio.no} \\
}

\renewcommand{\headeright}{}
\renewcommand{\undertitle}{}
\renewcommand{\shorttitle}{K\i z\i laslan and Vitelli}

\begin{document}
\maketitle

\begin{abstract}

In recent years, mixture cure models have gained increasing popularity in survival analysis as an alternative to the Cox proportional hazards model, particularly in settings where a subset of patients is considered cured. The proportional hazards mixture cure model is especially advantageous when the presence of a cured fraction can be reasonably assumed, providing a more accurate representation of long-term survival dynamics. In this study, we propose a novel hierarchical Bayesian framework for the semiparametric mixture cure model, which accommodates both the inclusion and exclusion of a frailty component, allowing for greater flexibility in capturing unobserved heterogeneity among patients.
Samples from the posterior distribution are obtained using a Markov chain Monte Carlo method, leveraging a hierarchical structure inspired by Bayesian Lasso. 
Comprehensive simulation studies are conducted across diverse scenarios to evaluate the performance and robustness of the proposed models. 
Bayesian model comparison and assessment are performed using various criteria. 
Finally, the proposed approaches are applied to two well-known datasets in the cure model literature: the E1690 melanoma trial and a colon cancer clinical trial.
\end{abstract}


\section{Introduction}

In survival analysis, a fundamental objective is to model time-to-occurrence of a specific event, such as disease progression or death, while accounting for the influence of covariates. Traditional survival models, such as the Cox proportional hazards (PH) model, assume that all individuals in a study will eventually experience the event of interest given sufficient follow-up time. 
However, advances in medical treatments and early detection methods have led to more and more situations in which it is reasonable to assume that a subset of patients may never experience the event, indicating the presence of a cured fraction. 
This has motivated the development of cure rate models, which distinguish between individuals who remain at risk and those who are effectively cured. These models have been widely applied in cancer research, where certain treatments can lead to complete remission, preventing disease recurrence. 
Given the increasing complexity of modern clinical data, developing robust statistical models that account for the presence of a cured fraction remains essential for accurate inference and decision-making in medical research.

The mixture cure model (\cite{boag1949maximum}, \cite{berkson1952survival}) and the promotion time cure model (\cite{yakovlev1996stochastic}) are the most widely used approaches in survival analysis for modeling the presence of a cured fraction.
A recent comprehensive textbook on cure models was authored by  \cite{peng2021cure}, while \cite{amico2018cure} provided an extensive literature review on cure regression models.
The literature on cure rate models has grown substantially in recent years, in both classical and Bayesian frameworks, reflecting the increasing interest in accurately modeling long-term survivors in survival analysis.

In survival analysis, it is often assumed that all individuals share the same risk of experiencing the event of interest. 
However, this assumption may not be entirely realistic and practical in many real-world settings. 
For example, in cancer studies, patients with similar clinical characteristics may still have different prognoses due to genetic or biological factors not recorded in the study. Such unobserved factors introduce heterogeneity that cannot be explained by the available covariates. Frailty models address this issue by incorporating a random effect -- named frailty -- that captures individual-specific variability in risk.
Incorporating frailty into cure models allows for a more flexible and realistic representation of the population, improving inference and prediction in the presence of latent differences among patients.
\cite{price2001modelling} first introduced a gamma frailty term into the latency distribution of mixture cure models, later extended by \cite{peng2008estimation}, and diverse approaches have been developed to incorporate frailty under different assumptions and modeling frameworks \cite{de2017bayesian, cancho2021bayesian, karamoozian2021bayesian, kizilaslan2025weibull}.
In a broader survival context, \cite{nipoti2018bayesian} proposed a Bayesian semiparametric partially proportional hazards frailty model for clustered time-to-event data, while \cite{choi2020joint} extended frailty modeling by introducing nested frailty models to jointly analyze recurrent and terminal events.

Another important methodological direction in survival and cure modeling focuses on flexible specifications of the baseline hazard function. The piecewise exponential (PE) model, which assumes a constant hazard within predefined time intervals, provides a semiparametric framework capable of approximating both monotonic and non-monotonic hazard shapes. 
This flexibility has motivated its adoption in a wide range of cure models, from promotion time cure models to more general cure rate frameworks \cite{chen2002bayesian, yin2005cure, lambert2007estimating, ibrahim2012bayesian, de2020bayesian, ibrahim2001bayesian}, particularly in the analysis of cancer survival data.

A Bayesian formulation of the promotion time cure model was first introduced by \cite{chen1999new}. 
\cite{ibrahim2001bayesian} proposed a hierarchical Bayesian semiparametric promotion time cure model using a PE baseline hazard with a smoothing parameter controlling tail parametricity.
Subsequently, Bayesian inference for cure rate models was comprehensively discussed in a dedicated section of the well-known Bayesian survival analysis book by \cite{ibrahim2001book}. 
\cite{chen2002bayesian} compared Bayesian parametric and semiparametric promotion time cure models, using the PE proportional hazards model as a benchmark.
Since then, numerous Bayesian studies have been published, with increasing attention in recent years due to both methodological and computational advances in Bayesian inferential approaches. In the current cure rate literature, Bayesian methods are predominantly tailored to the promotion time cure model and related approaches.  
\cite{yin2005bayesian} was the first to investigate the promotion time cure frailty model within a Bayesian framework, employing a semiparametric structure with a PE baseline distribution.
\cite{yin2005cure} proposed a general class of cure rate models by applying a Box–Cox transformation to the population survival function and proposed a Bayesian semiparametric model with a PE baseline survival function.
\cite{ibrahim2012bayesian} conducted a Bayesian analysis of the Eastern Cooperative Oncology Group (ECOG) phase III clinical trial melanoma datasets, employing a promotion time cure model framework.
\cite{demarqui2014fully} proposed a fully semiparametric Bayesian approach for the promotion time cure model of \cite{chen1999new} and \cite{ibrahim2001bayesian}, modeling the baseline hazard via a PE specification with a random time partition. \cite{psioda2018bayesian} developed a general Bayesian clinical trial design methodology that incorporates historical data through a promotion time cure model. Finally, \cite{barriga2019new} and \cite{cancho2021bayesian} extended the promotion time cure model by incorporating a frailty term to capture unobserved heterogeneity, the former via classical likelihood-based inference, while the latter via a Bayesian approach based on Markov chain Monte Carlo (MCMC), with both studies illustrating their methods using a colorectal cancer dataset.

Despite earlier work on mixture cure models dating back to \cite{yin2005cure} around $2005$, only in recent years several Bayesian approaches to cure models have been proposed. 
\cite{de2020bayesian} introduced a competing-risks cure rate model based on the power PE distribution, and \cite{lazaro2020approximate} investigated approximate Bayesian inference for mixture cure models using the integrated nested Laplace approximation (INLA); both applied their methods to the ECOG melanoma clinical trial dataset.
\cite{karamoozian2021bayesian} proposed two parametric Bayesian mixture cure frailty models with discrete and continuous frailty distributions, and applied both to the analysis of a gastric cancer dataset.
\cite{gressani2022laplacian} further advanced approximate Bayesian inference for mixture cure models by combining B-spline approximations of the baseline survival function with a Laplace approximation, differing from the INLA approach of \cite{lazaro2020approximate}.
\cite{wang2022bayesian} developed a Bayesian accelerated failure time mixture cure model, where the unknown error distribution of the latency component was flexibly modeled through a Dirichlet process.
Most recently, \cite{papastamoulis2025bayesian} proposed a fully Bayesian estimation framework for a broad class of cure rate models under the assumption that the promotion time follows a Weibull distribution.

In this study, we propose Bayesian hierarchical modeling frameworks for the semiparametric mixture cure (SMC) and semiparametric mixture cure frailty (SMCF) models that are suited to the case of very high-dimensional covariates. Therefore, the emphasis of our proposal is both on methodological developments, but also on performance assessment and comparative analysis as compared to the above mentioned Bayesian literature. 
\cite{lee2011bayesian} first proposed a Bayesian hierarchical model with a lasso-type shrinkage prior based on the scale-mixture representation of the Laplace distribution for variable selection in high-dimensional survival data. We extend this idea by incorporating the shrinkage prior into both the incidence and latency components of the model, thus not only enhancing parameter estimation, but also providing a foundation for future extensions to Bayesian variable selection.
We formulate a novel hierarchical structure and derive a tailored MCMC algorithm for inference in both models. 
To benchmark our approach, we implement four alternative specifications in Stan \citeyear{Stan} via the R interface, \texttt{RStan} \citeyear{Rstan}, and systematically compare their performance with our MCMC method.
An extensive simulation study under diverse scenarios is then carried out, providing a comprehensive evaluation of the proposed methodologies relative to existing alternatives. 
To demonstrate practical utility, we apply the proposed models to two benchmark datasets -- the E1690 melanoma trial and a colon cancer study -- showing their ability to capture cure fractions and covariate effects in clinically relevant survival settings.

This work extends prior research in several directions. Although our model is encompassed by the framework of \cite{yin2005cure}, we develop a distinct Bayesian inference strategy supported by more extensive simulation studies. Compared with \cite{lazaro2020approximate}, our fully Bayesian approach provides an alternative to approximate inference via INLA. In contrast to \cite{karamoozian2021bayesian}, our frailty model allows separate regression structures in the incidence and latency components, offering greater modeling flexibility.

The rest of this article is organized as follows: in Section \ref{sec:SMCM}, we describe the SMC model (SMCM) and develop our Bayesian hierarchical inference approach, detailing the proposed MCMC algorithm, along with two alternative Bayesian specifications implemented in \texttt{RStan}.
In Section \ref{sec:SMCFM}, we develop our novel methods for the SMCF model (SMCFM), extending the framework of the previous section, and we also introduce Bayesian model assessment criteria.
In Section \ref{sec:simulation}, we conduct a comprehensive Monte Carlo simulation study to evaluate the performance of our proposed MCMC methods for both the SMCM and the SMCFM, benchmarking them against four alternative implementations developed in \texttt{RStan}.
In Section \ref{sec:data_analysis}, we illustrate the practical utility of the proposed methodologies by analyzing two benchmark datasets, the E1690 melanoma and colon cancer studies, and discuss and compare the results with those obtained from existing models.
Finally, Section \ref{sec:conclusion} concludes the article with a summary of findings and a discussion of possible directions for future research.

\section{The Semiparametric Mixture Cure Model}\label{sec:SMCM}

Let the random variable $T$ represent the lifetime of interest with 
survival function denoted by $S_{pop}(t), t\in[0,+\infty)$. Let $Y$ be the indicator for a subject eventually $(Y=1)$ or never $(Y=0)$ experiencing the event of interest, with $\pi =P(Y=1)$ representing the probability of a subject being susceptible (or uncured) for the event of interest. Among the subjects for whom $Y=0$, the survival function is $S(t|Y=0)=1, \forall t\in[0,+\infty)$, and for those who experience the event ($Y=1$), the survival function and the probability density function are $S(t|Y=1)$ and $f(t|Y=1)$, respectively. $Y$ is not observed for a censored subject. The population survival function is therefore defined as
\begin{equation}
S_{pop}(t)=1-\pi +\pi S(t | Y=1).  \label{S_pop0}
\end{equation}%
Note that since $S_{pop}(t)\rightarrow 1-\pi $ as $t\rightarrow +\infty $, $S_{pop}(t)$ is not a proper survival function. The uncured rate $\pi $ and the survival function of the uncured subjects $S(t | Y=1)$ are also referred to as the incidence and the latency distribution, respectively.

The basic model introduced in (\ref{S_pop0}) can be extended to include the covariates associated with
the incidence and latency distributions. Let us denote by $\mathbf{x}$ and $\mathbf{z}$ the covariates that have an effect on the latency
distribution and incidence, respectively. Then, model (\ref{S_pop0}) can be rewritten as

\begin{equation}
S_{pop}(t|\mathbf{x,z})=1-\pi (\mathbf{z})+\pi (\mathbf{z})S(t|Y=1,\mathbf{x}%
),  \label{S_pop1}
\end{equation}%
where $\pi (\mathbf{z})$ is the probability of a subject being uncured
conditionally on $\mathbf{z}$, and $S(t|Y=1,\mathbf{x})$ is the survival
function of the lifetime distribution of uncured subjects conditionally on $%
\mathbf{x}$. 
Concerning the modeling of the effect of the covariates $\mathbf{z}$ on the incidence, as previously proposed in \cite{farewell1982use} 
we use a logistic regression model of the form $\pi(\mathbf{z}) = e^{\mathbf{z}^\top \mathbf{b}}/(1+e^{\mathbf{z}^\top \mathbf{b}}),$ where $\mathbf{z}^\top \in\mathbb{R}^{n \times p_1+1} $ is a covariate matrix, with columns $\mathbf{z}_1, \cdots, \mathbf{z}_n \in\mathbb{R}^{p_1+1}$, and $\mathbf{b}= (b_0,b_1, \cdots, b_{p_1})^\top \in\mathbb{R}^{p_1+1} $ is a vector of unknown regression coefficients. When the mixture cure model defined in (\ref{S_pop1}) is specified via proportional hazards, we get the following PH mixture cure model.

\begin{equation}
S_{pop}(t|\mathbf{x,z})=1-\pi (\mathbf{z})+\pi (\mathbf{z}) S_{0}(t)^{\exp(\mathbf{x}^\top \bm{\beta} )},  \label{S_pop2}
\end{equation}%
where $S_{0}(t)$ is the baseline survival function,  $\mathbf{x}^\top \in\mathbb{R}^{n \times p_2}$ is the covariate matrix, with columns $\mathbf{x_1}, \cdots, \mathbf{x_n} \in\mathbb{R}^{p_2}$, and $\bm{\beta} = (\beta_1, \cdots, \beta_{p_{2}})^ \top \in\mathbb{R}^{p_2}$ is the vector of unknown regression coefficients for the latency distribution.

To flexibly characterize the baseline hazard, we model $S_0(t)$ using the piecewise exponential distribution as in the cure model studies of \cite{chen2002bayesian, yin2005cure, ibrahim2012bayesian, de2020bayesian}.
Let $t_{i}$ be the observed survival time for the $i$th subject, $i=1,\cdots,n$, and $\delta_{i}$ an indicator function for censoring, with $\delta_{i}=1$ for the uncensored subject and $\delta_{i}=0$ for the censored subject. We construct a finite partition of the time axis, $s_1=0 <s_2<\cdots< s_{J+1}$, with $s_{J+1}>t_i$ for all $i=1,\cdots,n$. We then have $J$ intervals $(s_1,s_2], (s_2,s_3], \cdots, (s_{J},s_{J+1}]$, and we assume a constant hazard $\lambda_j,\; j=1,\cdots,J$ for each interval. Note that this setup reduces to a parametric exponential model when $J=1$. 

Let the observed data be $\mathbf{D}_{i}=(\delta _{i},t_{i},\mathbf{x}_{i},\mathbf{z}_{i} ),$ $i=1,...,n,$ and $\bm{\lambda}=(\lambda_1,\cdots, \lambda_J)^\top $ be a vector of constant hazards.
Also, $\mathbf{z}_{i}\in\mathbb{R}^{p_1}$ and $\mathbf{x}_{i}\in\mathbb{R}^{p_2}$ are respectively observed covariates associated to the cure and the survival parts of the model for the $i$th subject, $i=1,...,n$. Then, the likelihood function of $(\mathbf{b},\bm{\beta}, \bm{\lambda})$ for the right-censored observed survival data $\mathbf{D}=(\mathbf{D}_{1},...,\mathbf{D}_{n})$ is 
\begin{eqnarray}
L(\mathbf{b}, \bm{\beta}, \bm{\lambda}|\mathbf{D})
&=&\prod_{i=1}^{n}  f_{pop}(t_{i}|\mathbf{x}_{i},\mathbf{z}_{i}, \bm{\lambda})^{\delta_{i}}  S_{pop}(t_{i}|\mathbf{x}_{i}\mathbf{,z}_{i},\bm{\lambda}) ^{1-\delta _{i}}   \notag
\label{l_observed}
\end{eqnarray}
where $f_{pop}(t|\mathbf{x}, \mathbf{z},\bm{\lambda} )$ and $S_{pop}(t|\mathbf{x}, \mathbf{z},\bm{\lambda} )$ are the population density and survival functions such that 
\begin{eqnarray*}
    f_{pop}(t|\mathbf{x},\mathbf{z},\bm{\lambda}) =  \pi(\mathbf{z})\; h_0(t|\bm\lambda) \exp (\mathbf{x}^\top \bm{\beta}) \; S(t|\mathbf{x},\bm{\lambda}),
\end{eqnarray*}
and 
\begin{eqnarray*}
    S_{pop}(t|\mathbf{x},\mathbf{z},\bm{\lambda}) = 1-\pi(\mathbf{z})  + \pi(\mathbf{z}) S(t|\mathbf{x},\bm{\lambda}),
\end{eqnarray*}
where 
\begin{eqnarray}
    S(t|\mathbf{x},\bm{\lambda}) &=& \exp \left \{ -\left ( \lambda_j(t-s_j) + \sum_{k=2}^{j} \lambda_{k-1}(s_{k}-s_{k-1}) \right ) \exp (\mathbf{x}^\top \bm{\beta}) \right \} \notag \\
    &=&  \exp \left \{ -H_0(t| \bm{\lambda}  ) \; \exp (\mathbf{x}^\top \bm{\beta}) \right \},
\end{eqnarray}
with $s_{j}<t \leq s_{j+1}$, $\mathbf{b}\in\mathbb{R}^{p_1+1}$ and $\bm{\beta}\in\mathbb{R}^{p_2}$ being vectors of unknown regression coefficients for the covariates $\mathbf{x}$ and $\mathbf{z}$, respectively, where we indicate $\mathbf{z}_{i}^\top \mathbf{b} = b_{0} + \mathbf{z}_{i}^\top \mathbf{b} $, $i=1,...,n$.
Here, $h_0(t|\bm\lambda)=\lambda_j$ and $H_0(t | \bm{\lambda} ) = \lambda_j(t-s_j) + \sum_{k=2}^{j} \lambda_{k-1}(s_{k}-s_{k-1})$ for $s_{j}<t \leq s_{j+1}$ are the hazard and cumulative hazard functions of the piecewise exponential distribution in our setting. Finally, the likelihood function of $(\mathbf{b},\bm{\beta}, \bm{\lambda})$ takes the form
\begin{eqnarray}
L(\mathbf{b},  \bm{\beta},  \bm{\lambda}|\mathbf{D})
&=& \prod_{i=1}^{n}  \left\{\pi(\mathbf{z_i}) h_0(t_i|\bm\lambda) \exp (\mathbf{x_i}^\top \bm{\beta}) \exp \left \{ - H_0(t_i|\bm{\lambda}) \exp (\mathbf{x_i}^\top \bm{\beta}) \right \} \right\} ^{\delta_{i}} \notag \\
&\times& \left [ 1-\pi(\mathbf{z_i})  + \pi(\mathbf{z_i}) \exp \left \{ -H_0(t_i|\bm{\lambda}) \exp (\mathbf{x_i}^\top \bm{\beta}) + \mathbf{z}_{i}^\top \mathbf{b} \right \} \right ] ^{1-\delta _{i}}   \notag \\
\notag \\ 
&=& \prod_{i=1}^{n} 
\frac{ (e^{ \mathbf{z}_{i}^\top \mathbf{b}} h_0(t_i|\bm\lambda)  \exp (\mathbf{x_i}^\top \bm{\beta}))^{\delta_i  } }{ \left( 1+e^{ \mathbf{z}_{i}^\top \mathbf{b} } \right) }
\exp \left \{ -H_0(t_i|\bm{\lambda}) \exp (\mathbf{x_i}^\top \bm{\beta})  \delta_{i} v_{ij} \right \} \notag \\
&\times& \left [ 1 + \exp \left\{ -H_0(t_i|\bm{\lambda}) \exp (\mathbf{x_i}^\top \bm{\beta}) + \mathbf{z}_{i}^\top \mathbf{b} \right \} \right ] ^{1-\delta_{i} } \label{lik_SMCM}.
\end{eqnarray}

\subsection{Bayesian hierarchical model specification and full-conditional posterior distributions}{\label{sec:BayesSMCM}}

In this section, we introduce our novel Bayesian hierarchical approach to the SMCM. We begin by presenting the model structure and prior specifications, inspired by the penalized semiparametric Bayesian Cox model of \cite{lee2011bayesian, lee2015survival}, which builds on the Bayesian Lasso framework of \cite{park2008bayesian}. 

We assume that the coefficients $\mathbf{b}, \bm{\beta}, \bm{\lambda}$ in the model are assigned with independent priors. 
We follow a similar setup as that of \cite{yin2005cure} for the piecewise exponential part of the model, i.e. the hazard rate parameters $\bm{\lambda}$. We assume that the components of $\bm{\lambda}$, $\lambda_j, \; j=1,\cdots, J$ have a gamma prior distribution with shape parameter $a$ and rate parameter $b$, denoted $\text{Gamma}(a, b)$.  

\cite{lee2011bayesian, lee2015survival} employed the Bayesian Lasso within the penalized semiparametric Bayesian Cox model to enable Bayesian variable selection in high-dimensional settings. Building on this idea, we adapt the hierarchical structure of the Bayesian Lasso to our regression coefficients $\mathbf{b}$ and $\bm{\beta}$. The hierarchical representation of our full model is then specified as follows:
\begin{align}
& \mathbf{b}| \tau_{1}^2,\cdots,\tau_{p_1+1}^2 \sim \mathcal{N}(\mathbf{0}_{p_1+1}, \sigma^2 \mathbf{D}_{\tau}),\;\;\; \mathbf{D}_{\tau}=diag(\tau_{1}^2,\cdots,\tau_{p_1+1}^2) \notag \\
& \tau_{1}^2,\cdots,\tau_{p_1+1}^2|\eta^2 \sim \prod_{j=1}^{p_1+1}\frac{\eta^2}{2} e^{-\eta^2 \; \tau_{j}^2 /2} d\tau_{j}^2 ,\;\;\; 
 \pi(\eta^2) = \frac{\delta_{1}^{r_1}}{\Gamma(r_1)} (\eta^2)^{r_1-1} e^{-\delta_1 \eta^2},  \;\;\; \pi(\sigma^2) = \frac{1}{\sigma^2} \notag \\
& \bm{\beta}| \tau_{1}^{*{2}},\cdots,\tau_{p_2}^{*2} \sim \mathcal{N}(\mathbf{0}_{p_2}, \sigma^{*2} \mathbf{D}_{\tau^*}),\;\;\; \mathbf{D}_{\tau^*}=diag(\tau_{1}^{*{2}},\cdots,\tau_{p_2}^{*2}) \label{mcmc_hiear} \\
& \tau_{1}^{*2},\cdots,\tau_{p_2}^{*2}|\eta^{*2} \sim \prod_{j=1}^{p_2}\frac{\eta^{*2}}{2} e^{-\eta^{*2} \; \tau_{j}^{*2} /2} d\tau_{j}^{*2} ,\;\;\; 
 \pi(\eta^{*2}) = \frac{\delta_{2}^{r_{2}}}{\Gamma(r_2)} (\eta^{*2})^{r_2-1} e^{-\delta_2 \eta^{*2}},  \;\;\; \pi(\sigma^{*2}) = \frac{1}{\sigma^{*2}} \notag \\
& \pi(\bm{\lambda}) = \prod_{j=1}^{J} \frac{b^a}{\Gamma(a)} \lambda_{j}^{a-1} e^{-\lambda_{j}b}, \;\;\; a,b>0 \notag
\end{align}

Then, the joint posterior distribution of our model can be written as
\begin{eqnarray}
\pi(\mathbf{b},  \bm{\beta},  \bm{\lambda}, \bm{\tau},\bm{\tau}^*, \sigma^2,  \sigma^{*2}, \eta^2,  \eta^{*2} |\mathbf{D}) 
&\propto& L( \mathbf{b}, \bm{\beta}, \bm{\lambda} |\mathbf{D})  \pi(\mathbf{b}|\bm{\tau},\sigma^2)  \pi(\bm{\beta}| \bm{\tau}^*,\sigma^{*2}) \pi(\bm{\tau}|\eta^2) \pi(\bm{\tau}^*|\eta^{*2}) \notag \\
&\times& \pi(\eta^2)  \pi(\eta^{*2}) \pi(\sigma^2) \pi(\sigma^{*2})  \pi(\bm{\lambda}) \notag \\
&=& L(\mathbf{b}, \bm{\beta}, \bm{\lambda} | \mathbf{D} ) \prod_{j=1}^{p_1+1} \pi(b_j|\tau_j,\sigma^2) \pi(\tau_j|\eta^2)  \pi(\eta^2) \pi(\sigma^2) \notag \\
&\times& \prod_{j=1}^{p_2} \pi(\beta_j|\tau_j^{*},\sigma^{*2})  \pi(\tau_j^*|\eta^{*2} )  \pi(\eta^{*2})    \pi(\sigma^{*2}) \;\pi(\bm{\lambda})  \notag  \\ 
&=& \prod_{i=1}^{n} \left( \frac{e^{\mathbf{z}_{i}^\top \mathbf{b}}}{1+e^{ \mathbf{z}_{i}^\top \mathbf{b}}} h_0(t_i|\bm\lambda)  e^{\mathbf{x_i}^\top \bm{\beta}}  \right)^{\delta_{i}}  \left( \frac{1}{1+e^{ \mathbf{z}_{i}^\top \mathbf{b} }} \right)^{1-\delta _{i} } \notag \\
&\times& \exp \left \{ -H_0(t_i| \bm{\lambda} ) \; e^{\mathbf{x_i}^\top \bm{\beta}}  \; \delta _{i}  \right \}    \notag  \\ 
&\times& \left [ 1 + \exp \left\{ -H_0(t_i| \bm{\lambda} ) \; e^{\mathbf{x_i}^\top \bm{\beta}} 
+ \mathbf{z}_{i}^\top \mathbf{b} \right \} \right ] ^{1-\delta _{i} } \notag \\
&\times& \exp{ \left( -\frac{1}{2\sigma^2}  \mathbf{b}^\top \bm{D_{\tau}^{-1}} \mathbf{b} \right) } \;
\prod_{j=1}^{p_1+1} \frac{\eta^2}{2} e^{-\eta^2 \; \tau_{j}^2 /2} d\tau_{j}^2  \notag \\
&\times& \frac{\delta_1^{r_1}}{\Gamma(r_1)} (\eta^2)^{r_1-1} e^{-\delta_1 \eta^2} \; \frac{1}{\sigma^2} \notag \\
&\times& \exp{ \left( -\frac{1}{2\sigma^{*2}}  \bm{\beta}^\top \bm{D_{\tau^*}^{-1}} \bm{\beta} \right) } 
\prod_{j=1}^{p_2}\frac{\eta^{*2}}{2} e^{-\eta^{*2} \; \tau_{j}^{*2} /2} d\tau_{j}^{*2} \notag \\
&\times& \frac{\delta_2^{r_2}}{\Gamma(r_2)} (\eta^{*2})^{r_2-1} e^{-\delta_2 \eta^{*2}} \; \frac{1}{\sigma^{*2}}  \;
\prod_{j=1}^{J} \frac{b^a}{\Gamma(a)} \lambda_{j}^{a-1} e^{-b \lambda_{j}}. \label{full_post_SMCM} 
\end{eqnarray}
To perform inference on the parameters, we derive the full conditional posterior distributions of the model parameters $\mathbf{b}, \; \bm{\beta}, \; \bm{\lambda}, \;\bm{\tau}, \;\bm{\tau}^*, \; \sigma^2, \; \sigma^{*2}, \; \eta^2$ and $\eta^{*2}$ from the joint posterior distribution given in (\ref{full_post_SMCM}). 
In what follows, we present the conditional posterior distributions of $\mathbf{b}$, $\bm{\beta}$, and $\bm{\lambda}$. As they do not follow standard distributional forms, we set up and MCMC algorithm where they are sampled using Metropolis–Hastings (see Section \ref{sec:SMC_MCMC}).

The full conditional posterior distribution for $b_k, \; k=1,\cdots,p_1+1$ is 
\begin{eqnarray}
\pi(b_k| \mathbf{b}^{(-k)}, \bm{\beta},  \bm{\lambda}, \bm{\tau}, \sigma^2, \eta^2 , \mathbf{D}) 
&\propto&   \exp{ \left( -\frac{1}{2\sigma^2}  \mathbf{b}^\top \bm{D_{\tau}^{-1}} \mathbf{b} \right) }  \prod_{i=1}^{n} \frac{e^{\mathbf{z}_{i}^\top \mathbf{b} \delta_i } } { (1+e^{ \mathbf{z}_{i}^\top \mathbf{b}}) }  
  \notag \\ 
&\times& \left [ 1 + \exp \left\{ -H_0(t_i| \bm{\lambda} ) \; e^{\mathbf{x_i}^\top \bm{\beta}} 
+ \mathbf{z}_{i}^\top \mathbf{b} \right \} \right ] ^{1-\delta _{i} }, \label{posterior_b_k}
\end{eqnarray}
where $\mathbf{b}^{(-k)}$ represents the $\mathbf{b}$ vector without the $k^\text{th}$ element. 
The full conditional posterior distribution for $\beta_k, \; k=1,\cdots,p_2$ is 
\begin{eqnarray}
\pi(\beta_k| \bm{\beta}^{(-k)}, \mathbf{b},  \bm{\lambda}, \bm{\tau}^*, \sigma^{*2}, \eta^{*2} , \mathbf{D})  
&\propto& \exp{ \left( -\frac{1}{2\sigma^{*2}}  \bm{\beta}^\top \bm{D_{\tau^*}^{-1}} \bm{\beta} \right) } 
\prod_{i=1}^{n} \exp \left \{ -H_0(t_i| \bm{\lambda} ) \;  e^{\mathbf{x_i}^\top \bm{\beta}}  \; \delta _{i} \right \}  \notag \\
&\times&  e^{\mathbf{x_i}^\top \bm{\beta} \delta_{i} } \; \left [ 1 + \exp \left\{ -H_0(t_i| \bm{\lambda} ) \; e^{\mathbf{x_i}^\top \bm{\beta}} + \mathbf{z}_{i}^\top \mathbf{b}  \right \} \right ] ^{ 1-\delta _{i} } . \label{posterior_beta_k} 
\end{eqnarray}
The full conditional posterior distribution for $\lambda_k, \; k=1,\cdots,J$ is 
\begin{eqnarray}
\pi(\lambda_k| \bm{\lambda}^{(-k)}, \mathbf{b}, \bm{\beta}, \mathbf{D}) &\propto& 
\prod_{i=1}^{n}  h_0(t_i| \bm{\lambda})^{\delta_{i}}  \exp \left \{ -H_0(t_i| \bm{\lambda}) \; e^{\mathbf{x_i}^\top \bm{\beta}} \delta_{i}  \right \}  \notag \\
&\times& \left [ 1 + \exp \left\{ -H_0(t_i| \bm{\lambda} ) \; e^{\mathbf{x_i}^\top \bm{\beta}} + \mathbf{z}_{i}^\top \mathbf{b}  \right \} \right ] ^{ 1-\delta _{i} } \lambda_k^{a-1} e^{-b \lambda_k}, \label{posterior_lambda_k}
\end{eqnarray}
where $\bm{\beta}^{(-k)}$ and $\bm{\lambda}^{(-k)}$ have the same interpretation as $\mathbf{b}^{(-k)}$. Similarly to the model specifications in \cite{lee2011bayesian}, the full conditional posterior distributions of the parameters $ \bm{\tau}, \bm{\tau}^*, \sigma^2,  \sigma^{*2}, \eta^2 $ and  $\eta^{*2}$ follow well-known distributions  as detailed below
\begin{align}
& \frac{1}{\tau_{k}^{2}}| \bm{\tau}^{(-k)}, \mathbf{b}, \bm{\lambda},  \sigma^2, \eta^2 , \mathbf{D} 
\sim \text{Inv-Gaussian} \left(  \frac{\eta \sigma}{\mid b_k \mid } , \; \eta^2 \right), \label{posterior_tau} \\ 
& \frac{1}{\tau_{k}^{*^{2}}} | \bm{\tau}^{*(-k)}, \bm{\beta}, \bm{\lambda}, \sigma^{*2}, \eta^{*2} , \mathbf{D} 
\sim \text{Inv-Gaussian} \left(  \frac{\eta^* \sigma^*}{\mid \beta_k \mid }, \; \eta^{*2} \right), \label{posterior_tau_star}  \\
& \sigma^2 |  \mathbf{b}, \bm{\lambda}, \bm{\tau},  \eta^2 , \mathbf{D} 
\sim \text{Inv-Gamma} \left( \frac{p_1+1}{2}, \; \sum_{j=1}^{p_1+1}\frac{b_j^2}{2\tau_j^2}  \right),  \label{posterior_sigma} \\
& \sigma^{*2} |  \bm{\beta}, \bm{\lambda}, \bm{\tau}^*,  \eta^{*2} , \mathbf{D} 
\sim \text{Inv-Gamma} \left( \frac{p_2}{2}, \;  \sum_{j=1}^{p_2}\frac{\beta_j^2}{2\tau_j^{*2}} \right), \label{posterior_sigma_star}\\
& \eta^2 |  \mathbf{b}, \bm{\lambda}, \bm{\tau}, \sigma^2, \mathbf{D} 
\sim \text{Gamma}  \left( p_1+1+r_1, \; \delta_1 + \sum_{j=1}^{p_1+1}\frac{\tau_j^2}{2}  \right), \label{posterior_eta} \\
& \eta^{*2} | \bm{\beta}, \bm{\lambda}, \bm{\tau}^*, \sigma^{*2}, \mathbf{D} 
\sim \text{Gamma}  \left( p_2+r_2, \; \delta_2 + \sum_{j=1}^{p_2}\frac{\tau_j^{*2}}{2}  \right). \label{posterior_eta_star}
\end{align}

\subsection{Computational approaches}

In this section, we first describe our proposed MCMC algorithm to perform inference on the model parameters, which combines a Metropolis–Hastings scheme with Gibbs sampling. Then, for comparison with our MCMC approach, we implement two alternative Bayesian specifications using \texttt{RStan} \citeyear{Rstan}.

\subsubsection{Metropolis-within-Gibbs MCMC method for the SMCM}{\label{sec:SMC_MCMC}}
The detailed MCMC algorithm for the SMCM is presented as follows.

\begin{itemize}
    \item \textbf{Initialization.} Set the initial values of all parameters: $\bm{b}^{(0)}, \; \bm{\beta}^{(0)},\; \bm \lambda^{(0)}, \; \bm{\tau}^{(0)}, \bm{\tau}^{*(0)},\sigma^{2(0)}, \; \sigma^{*2(0)}, \eta^{2(0)}$ and $ \eta^{*2(0)} $. The index $g$ denotes the $g^\text{th}$ iteration, with $M$ being the total number of posterior samples. Set $g=1$.
    
    \item \textbf{Step 1.} Update $\bm{b}$ by performing a Metropolis–Hastings step based on the full conditional posterior of $b_k, \; k=1,\cdots, p_1+1 $ given in (\ref{posterior_b_k}), as detailed below
    \begin{itemize}
        \item [(i)] Set $k=1$.
        \item [(ii)] Sample a proposal $b_k^{(prop)}$ from a proposal distribution, $J_{g}(b_k^{(prop)} | b_k^{(g-1)}) = \mathcal{N}(b_k^{(prop)} | \mu_{b_k}^{(g-1)}, v_{b_k}^{(g-1)} )$, and $u_k$ from the uniform distribution on $(0,1)$, $u_k \sim \mathcal{U}(0,1) $. Note that the proposal variances are fixed at $v_{b_k}^{(g-1)} = 1$ as in \cite{zucknick2015nonidentical}.
        \item [(iii)] Calculate the acceptance probability:
        \begin{align}
            p_k = \frac{\pi( b_{k}^{(prop)} | \mathbf{b}^{(-k)(g-1)}, \bm{\beta}^{(g-1)},  \bm{\lambda}^{(g-1)}, \bm{\tau}^{(g-1)}, \sigma^{2 (g-1)}, \eta^{2 (g-1)} , \mathbf{D}) \; / \; J_{g}(b_k^{(prop)} | b_k^{(g-1)})} 
            {\pi( b_{k}^{(g-1)} | \mathbf{b}^{(-k)(g-1)}, \bm{\beta}^{(g-1)},  \bm{\lambda}^{(g-1)}, \bm{\tau}^{(g-1)}, \sigma^{2 (g-1)}, \eta^{2 (g-1)} , \mathbf{D}) \; / \; J_{g}(b_k^{(g-1)} | b_k^{(prop)})} 
        \end{align}
        \item[(iv)] If $u_k < p_k$ then $b_k^{(g)}=b_{k}^{(prop)}$ else $ b_k^{(g)}= b_k^{(g-1)}$. 
        \item[(v)] Stop if $k = p_1+1$. Otherwise set $k = k +1 $, and go to step (ii).
    \end{itemize}
    
        \item \textbf{Step 2.} Update $\bm{\beta}$ by performing a Metropolis–Hastings step based on the full conditional posterior of $\bm{\beta}_k, \; k=1,\cdots, p_2 $ given in (\ref{posterior_beta_k}), as detailed below
    \begin{itemize}
        \item [(i)] Set $k=1$.
        \item [(ii)] Sample a proposal $\beta_k^{(prop)}$ from a proposal distribution, $J_{g}(\beta_k^{(prop)} | \beta_k^{(g-1)}) = \mathcal{N}(\beta_k^{(prop)} | \mu_{\beta_k}^{(g-1)}, v_{\beta_k}^{(g-1)} )$ with the proposal variances are fixed at $v_{\beta_k}^{(g-1)} = 1$ and $u_k \sim \mathcal{U}(0,1)$.
        \item [(iii)] Calculate the acceptance probability:
        \begin{align}
            p_k = \frac{\pi( \beta_{k}^{(prop)} | \bm{\beta}^{(-k)(g-1)}, \bm{b}^{(g-1)},  \bm{\lambda}^{(g-1)}, \bm{\tau}^{*(g-1)}, \sigma^{*2 (g-1)}, \eta^{*2 (g-1)} , \mathbf{D}) \; / \; J_{g}(\beta_k^{(prop)} | \beta_k^{(g-1)})} 
            {\pi( \beta_{k}^{(g-1)} | \bm{\beta}^{(-k)(g-1)}, \bm{b}^{(g-1)},  \bm{\lambda}^{(g-1)}, \bm{\tau}^{*(g-1)}, \sigma^{*2 (g-1)}, \eta^{*2 (g-1)} , \mathbf{D}) \; / \; J_{g}(\beta_k^{(g-1)} | \beta_k^{(prop)})} 
        \end{align}
        \item[(iv)]  If $u_k < p_k$ then $\beta_k^{(g)}=\beta_{k}^{(prop)}$ else $\beta_k^{(g)}= \beta_k^{(g-1)}$.
        \item[(v)] Stop if $k = p_2$. Otherwise set $k = k +1 $, and go to step (ii).
    \end{itemize}

         \item \textbf{Step 3.} Update $\bm{\lambda}$ by performing a Metropolis–Hastings step based on the full conditional posterior of $\bm{\lambda}_k, \; k=1,\cdots, J $ given in (\ref{posterior_lambda_k}), as detailed below
    \begin{itemize}
        \item [(i)] Set $k=1$.
        \item [(ii)] Sample a proposal $\lambda_k^{(prop)}$ from a proposal distribution, $J_{g}(\lambda_k^{(prop)} | \lambda_k^{(g-1)}) = \text{Gamma}(\beta_k^{(prop)} | a_{\lambda_k}^{(g-1)}, \\ b_{\lambda_k}^{(g-1)} )$ with the rate parameters are fixed at $b_{\lambda_k}^{(g-1)}= 1$ and $u_k \sim \mathcal{U}(0,1)$.
        \item [(iii)] Calculate the acceptance probability:
        \begin{align}
            p_k &= \frac{\pi( \lambda_{k}^{(prop)} | \bm{\lambda}^{(-k)(g-1)}, \bm{b}^{(g-1)}, \bm{\beta}^{(g-1)}, \mathbf{D})  / J_{g}(\lambda_k^{(prop)} | \lambda_k^{(g-1)}) } 
            {\pi( \lambda_{k}^{(g-1)} | \bm{\lambda}^{(-k)(g-1)}, \bm{b}^{(g-1)}, \bm{\beta}^{(g-1)},  \mathbf{D}) / J_{g}(\lambda_k^{(g-1)} | \lambda_k^{(prop)}) } 
        \end{align}
        \item[(iv)]  If $u_k < p_k$ then $\lambda_k^{(g)}=\lambda_{k}^{(prop)}$ else $\lambda_k^{(g)}= \lambda_k^{(g-1)}$.
        \item[(v)] Stop if $k = J$. Otherwise set $k = k +1 $, and go to step (ii).
    \end{itemize}
    
    \item \textbf{Step 4.} Sample $1/\tau_{k}^2, \; k=1, \cdots, p_1$ from its full conditional distribution given in (\ref{posterior_tau}).
    \item \textbf{Step 5.} Sample $1/\tau_{k}^{*2}, \; k=1, \cdots, p_2$ from its full conditional distribution given in (\ref{posterior_tau_star}).
    
    \item \textbf{Step 6.} Sample $\sigma^2 $ from its full conditional distribution given in (\ref{posterior_sigma}).
    \item \textbf{Step 7.} Sample $\sigma^{*2} $ from its full conditional distribution given in (\ref{posterior_sigma_star}).
    
    \item \textbf{Step 8.} Sample $\eta^2 $ from its full conditional distribution given in (\ref{posterior_eta}).
    \item \textbf{Step 9.} Sample $\eta^{*2} $ from its full conditional distribution given in (\ref{posterior_eta_star}).

    \item \textbf{Stopping criterion.} If $ g= M$, stop. Otherwise, set $g= g+ 1$ and go to \textbf{Step} $\mathbf{1}$.     
\end{itemize}

\subsubsection{Alternative Bayesian specifications via \texttt{RStan}} {\label{sec:SMC_Rstan}}

For comparison with our proposed Bayesian approach where inference is performed via an MCMC algorithm, we also consider two alternative specifications implemented using the \texttt{RStan} package: (i) the hierarchical semiparametric mixture cure model (HSMCM) structure given in (\ref{mcmc_hiear}), and (ii) a regular specification employing normal priors for the regression coefficients, following the approach of \cite{yin2005cure}. Specifically, the HSMCM implementation in \texttt{RStan}, which is called as \textit{HSMCM(RStan)}, works under the following prior assumptions:
\begin{align}
& \lambda_j \sim \text{Gamma}(a,b), \;\;\; j=1,\cdots,J \notag \\
& b_j \mid \tau_{j}^2, \sigma^2 \sim \mathcal{N}(0, \sigma^2 \tau_{j}^2), \;\;\;
 \tau_{j}^2 \mid \eta^2 \sim \text{Exponential}(\eta^2/2), \;\;\; j=1,\cdots,p_1+1 \notag \\
& \eta^2 \sim \text{Gamma}( r_1, \delta_{1}), \;\;\; \sigma^2  \sim \mathcal{U}(0, 1000)  \notag \\
& \beta_j \mid \tau_{j}^{*{2}}, \sigma^{*2} \sim \mathcal{N}(0, \sigma^{*2} \tau_j^{*{2}}), \;\;\;  
 \tau_{j}^{*{2}} \mid \eta^{*{2}} \sim \text{Exponential}(\eta^{*{2}}/2), \;\;\; j=1,\cdots,p_2  \notag \\
 & \eta^{*{2}} \sim \text{Gamma}( r_2, \delta_{2}), \;\;\; \sigma^{*2}  \sim \mathcal{U}(0, 1000).  \notag 
\end{align} 
In the second specification, called \textit{SMCM(RStan)}, normal priors are assigned to the regression coefficients as in \cite{yin2005cure}, namely
$\mathbf{b} \sim N(\mathbf{0}, \bm{\Sigma_{b}})$ and $\bm{\beta} \sim N(\mathbf{0}, \bm{\Sigma_{\beta}})$  with the gamma prior for $\bm \lambda$, i.e. $\bm\lambda \sim \text{Gamma}(a,b)$.

\section{The Semiparametric Mixture Cure Frailty Model (SMCFM)} \label{sec:SMCFM}

Frailty models, widely used in medical and epidemiological research, extend proportional hazards models by incorporating an unobserved random effect to better capture the heterogeneity of survival outcomes. The frailty term enters multiplicatively into the hazard function, thereby accounting for unmeasured heterogeneity among individuals.
Let $W_i$ be a non-negative frailty random variable associated to the $i$th subject, with cumulative distribution function $F_{W_i}(w)$.
The hazard function of the $i$th subject with frailty $W_i$ is then 
$h(t|W_i) = W_i h_{0}(t) \exp (\mathbf{x_i}^\top \bm{\beta}),$
where $h_{0}(t)$ is a baseline hazard function common for all subjects and $\mathbf{x_i}$ is a covariate vector for the $i$th subject. If we include the frailty in the latency distribution in model (\ref{S_pop2}), the conditional survival function given the frailty $W$ takes the form $S(t|Y=1,W,\mathbf{x})=\exp (-W e^{ \mathbf{x}^\top \bm{\beta}} H_{0}(t)),$
where $H_{0}(t)$ is the baseline cumulative hazard function. Then, the marginal survival function of uncured subjects based on the frailty model is
given by $ S(t | Y=1,\mathbf{x}) = L_{W}(e^{\mathbf{x}^\top \bm{\beta}} H_{0}(t)),$
where $L_{W}(s)=E(e^{-ws})$ is the Laplace transformation of the frailty distribution $W$.
The PH mixture cure model in (\ref{S_pop2}) is extended 
by incorporating a frailty term $W,$ leading to the following formulation:
\begin{equation}
S_{pop}(t | \mathbf{x,z})=1-\pi (\mathbf{z})+\pi (\mathbf{z})L_{W}(e^{\mathbf{x}^\top \bm{\beta}} H_{0}(t)). \label{S_pop_MCFM}
\end{equation}%
This model was first introduced by \cite{peng2008estimation} and called the mixture cure frailty model (MCFM). It reduces to the PH mixture cure model in (\ref{S_pop2}) when there is no frailty effect, namely $W\equiv 1$, and it reduces to a standard frailty model when there is no cure fraction existing in the population, namely $\pi (\mathbf{z})\equiv 1$.

The Gamma distribution offers considerable flexibility, allowing it to accommodate both increasing and decreasing hazard rates. Its closed-form and straightforward Laplace transform make it a convenient choice for modeling unobserved heterogeneity. Consequently, it is widely employed as a frailty distribution in survival analysis and mixture cure models, as highlighted in \cite{price2001modelling} and \cite{peng2008estimation}.
As noted in \cite{peng2008identifiability}, mixture cure frailty models face identifiability challenges, which require fixing the mean of the frailty distribution at $1$.
We assume that the frailty $W$ follows a gamma distribution
with mean $1$ and variance $1/\theta $. The Laplace transformation of the frailty is then $L_{w}(s)=(1+s/\theta )^{-\theta }$.
Then, the survival function of model (\ref{S_pop_MCFM}) becomes
\begin{equation}
S_{pop}(t | \mathbf{x,z})=1-\pi (\mathbf{z})+\pi (\mathbf{z}) 
\left( 1 + \frac{ e^{\mathbf{x}^\top \bm{\beta}} \; H_{0}(t) }{\theta} \right) ^{-\theta}.
\end{equation}

As in the previous section, we model the baseline cumulative hazard function $H_0(t)$ using a piecewise exponential distribution.
Let the observed data be $\mathbf{D}_{i}=(\delta _{i},t_{i},\mathbf{x}_{i},\mathbf{z}_{i} ),$ $i=1,...,n,$ and $\bm{\lambda}=(\lambda_1,\cdots, \lambda_J)^\top$ be a vector of constant hazards.
Also $\mathbf{z}_{i}\in\mathbb{R}^{p_1+1}$ and $\mathbf{x}_{i}\in\mathbb{R}^{p_2}$ are respectively observed covariates associated to the cure and the survival parts of the model for the $i$th subject, $i=1,...,n$. Then, the likelihood function of $(\mathbf{b},\bm{\beta}, \bm{\lambda}, \theta)$ for the right-censored observed survival data $\mathbf{D}=(\mathbf{D}_{1},...,\mathbf{D}_{n})$ is obtained by plugging in the following population density and survival functions 
\begin{eqnarray*}
    f_{pop} (t | \mathbf{x}, \mathbf{z}, \bm{\lambda}) &= & \pi(\mathbf{z})\; h_0(t | \bm{\lambda} ) \exp (\mathbf{x}^\top \bm{\beta}) \; 
    \left( 1 + \frac{ e^{\mathbf{x}^\top \bm{\beta}} \; H_0(t | \bm{\lambda} ) }{\theta} \right) ^{-\theta-1}\\
    S_{pop}(t | \mathbf{x}, \mathbf{z}, \bm{\lambda}) &= &1-\pi(\mathbf{z})  + \pi(\mathbf{z}) 
        \left( 1 + \frac{ e^{\mathbf{x}^\top \bm{\beta}} \; H_0(t | \bm{\lambda} ) }{\theta} \right) ^{-\theta}
\end{eqnarray*}
into the likelihood function
\begin{eqnarray}
L(\mathbf{b}, \bm{\beta}, \bm{\lambda}, \theta  |\mathbf{D})
&=&\prod_{i=1}^{n} f_{pop}(t_{i} | \mathbf{x}_{i}\mathbf{,z}_{i}) ^{\delta_{i}}  \; S_{pop}(t_{i} | \mathbf{x}_{i}\mathbf{,z}_{i}) ^{1-\delta _{i} },  \label{lik_SMCFM}
\end{eqnarray}
where $h_0(t | \bm{\lambda} )$ and $H_0(t  | \bm{\lambda} )$ are the hazard and cumulative hazard functions of the piecewise exponential distribution.

\subsection{Bayesian inference in the SMCFM} \label{sec:BayesSMCFM}

In this section, we extend the hierarchical framework introduced for the SMCM in Section \ref{sec:BayesSMCM} to the SMCFM. We adopt the same Bayesian hierarchical structure given in (\ref{mcmc_hiear}) for the incidence and latency components, while additionally incorporating a gamma-distributed frailty term $W$ to account for unobserved heterogeneity. 
A non-informative prior, such as a gamma distribution with large variance, is assigned to the frailty parameter $\theta$. 
Given the similarities with the SMCM, we briefly summarize the MCMC algorithm, highlighting the modifications required to accommodate the frailty term. 

The joint posterior distribution of all model parameters in the SMCFM is given as
\begin{eqnarray}
\pi(\mathbf{b},  \bm{\beta},  \bm{\lambda}, \theta,  \bm{\tau},\bm{\tau}^*, \sigma^2,  \sigma^{*2}, \eta^2,  \eta^{*2} | \mathbf{D}) 
&=& L(\mathbf{b}, \bm{\beta}, \bm{\lambda}, \theta | \mathbf{D} ) \prod_{j=1}^{p_1+1} \pi(b_j \mid \tau_j,\sigma^2) \pi(\tau_j \mid \eta^2)  \pi(\eta^2) \pi(\sigma^2) \notag \\
&\times& \prod_{j=1}^{p_2} \pi(\beta_j \mid \tau_j^{*},\sigma^{*2})  \pi(\tau_j^* \mid \eta^{*2} )  \pi(\eta^{*2})  \pi(\sigma^{*2}) \;\pi(\bm{\lambda}) \pi(\theta)  \notag  \\ 
&=& \prod_{i=1}^{n}  \left( \frac{e^{\mathbf{z}_{i}^\top \mathbf{b}}}{1+e^{ \mathbf{z}_{i}^\top \mathbf{b}}} h_0(t_i| \bm{\lambda} )\;   e^{\mathbf{x_i}^\top \bm{\beta}}  \; \left( 1 + \frac{ e^{\mathbf{x_i}^\top \bm{\beta}} \; H_0(t_i | \bm{\lambda} ) }{\theta} \right) ^{-\theta-1}  \right)^{\delta _{i} }  \notag \\
&\times& \left( \frac{1}{1+e^{ \mathbf{z}_{i}^\top \mathbf{b} }} \right)^{ 1-\delta _{i} }
\left[ 1 + e^{ \mathbf{z}_{i}^\top \mathbf{b} } \left( 1 + \frac{ e^{\mathbf{x_i}^\top \bm{\beta}} \; H_0(t_i | \bm{\lambda} ) }{\theta} \right) ^{-\theta}  \right] ^{ 1-\delta _{i} }\notag \\
&\times& \exp{ \left( -\frac{1}{2\sigma^2}  \mathbf{b}^\top \bm{D_{\tau}^{-1}} \mathbf{b} \right) } \;
\prod_{j=1}^{p_1+1} \frac{\eta^2}{2} e^{-\eta^2 \; \tau_{j}^2 /2} d\tau_{j}^2  \notag \\
&\times& \frac{\delta_1^{r_1}}{\Gamma(r_1)} (\eta^2)^{r_1-1} e^{-\delta_1 \eta^2} \; \frac{1}{\sigma^2} \notag \\
&\times&  \exp{ \left( -\frac{1}{2\sigma^{*2}}  \bm{\beta}^\top \bm{D_{\tau^*}^{-1}} \bm{\beta} \right) } \;
\prod_{j=1}^{p_2}\frac{\eta^{*2}}{2} e^{-\eta^{*2} \; \tau_{j}^{*2} /2} d\tau_{j}^{*2} \notag \\
&\times& \frac{\delta_2^{r_2}}{\Gamma(r_2)} (\eta^{*2})^{r_2-1} e^{-\delta_2 \eta^{*2}} \; \frac{1}{\sigma^{*2}} \notag \\
&\times&  \prod_{j=1}^{J} \frac{b^a}{\Gamma(a)} \lambda_{j}^{a-1} e^{-b \lambda_{j}} 
\; \frac{d^c}{\Gamma(c)} \theta^{c-1} e^{-d \theta }.\label{full_post_SMCFM} 
\end{eqnarray}

We derive the full conditional posterior distributions of the model parameters $\mathbf{b}, \; \bm{\beta}, \; \bm{\lambda}, \; \theta, \; \bm{\tau}, \; \bm{\tau}^*, \sigma^2,  \; \sigma^{*2},\;  \eta^2$ and $\eta^{*2}$ from the joint posterior distribution given in (\ref{full_post_SMCFM}).
As for the SMCM (Section \ref{sec:BayesSMCM}), the conditional posterior distributions of $\mathbf{b}$, $\bm{\beta}$, $\bm{\lambda}$, and $\theta$ are sampled using a Metropolis–Hastings scheme.

The full conditional posterior distribution for $b_k, \; k=1,\cdots,p_1+1$ is 
\begin{eqnarray}
\pi(b_k \mid \mathbf{b}^{(-k)}, \bm{\beta},  \bm{\lambda}, \theta, \bm{\tau}, \sigma^2, \eta^2 , \mathbf{D}) &\propto& 
 \exp{ \left( -\frac{1}{2\sigma^2}  \mathbf{b}^\top \bm{D_{\tau}^{-1}} \mathbf{b} \right) } \; 
\prod_{i=1}^{n} \prod_{j=1}^{J}  \frac{e^{\mathbf{z}_{i}^\top \mathbf{b} \delta_i } } { (1+e^{ \mathbf{z}_{i}^\top \mathbf{b}}) }  \notag \\
&\times& \left[ 1 + e^{ \mathbf{z}_{i}^\top \mathbf{b} } \left( 1 + \frac{ e^{\mathbf{x_i}^\top \bm{\beta}} \; H_0(t_i | \bm{\lambda} ) }{\theta} \right) ^{-\theta}  \right] ^{ 1-\delta _{i} }. \notag \\ \label{posterior_bk_SMCFM}
\end{eqnarray} 
The full conditional posterior distribution for $\beta_k, \; k=1,\cdots,p_2$ is 
\begin{eqnarray}
\pi(\beta_k \mid \bm{\beta}^{(-k)}, \mathbf{b}, \bm{\lambda}, \theta,  \bm{\tau}^*, \sigma^{*2}, \eta^{*2} , \mathbf{D}) &\propto& 
\exp{ \left( -\frac{1}{2\sigma^{*2}}  \bm{\beta}^\top \bm{D_{\tau^*}^{-1}} \bm{\beta} \right) } 
\prod_{i=1}^{n}  
\left( 1 + \frac{ e^{\mathbf{x_i}^\top \bm{\beta}} \; H_0(t_i | \bm{\lambda} ) }{\theta} \right) ^{-(\theta+1) \delta _{i} } \notag \\
&\times& e^{\mathbf{x_i}^\top \bm{\beta} \delta_i } \;  \left[ 1 + e^{ \mathbf{z}_{i}^\top \mathbf{b} } \left( 1 + \frac{ e^{\mathbf{x_i}^\top \bm{\beta}} \; H_0(t_i | \bm{\lambda} ) }{\theta} \right) ^{-\theta}  \right] ^{ 1-\delta _{i} } .
\label{posterior_betak_SMCFM}
\end{eqnarray}
The full conditional posterior distribution for $\lambda_k, \; k=1,\cdots,J$ is 
\begin{eqnarray}
\pi(\lambda_k | \bm{\lambda}^{(-k)}, \mathbf{b}, \bm{\beta}, \theta, \mathbf{D}) &\propto& 
\prod_{i=1}^{n} 
h_0(t_i | \bm{\lambda} )^{\delta _{i}}  \left( 1 + \frac{ e^{\mathbf{x_i}^\top \bm{\beta}} \; H_0(t_i | \bm{\lambda} ) }{\theta} \right) ^{-(\theta+1) \delta _{i}  } \notag \\
&\times& \left[ 1 + e^{ \mathbf{z}_{i}^\top \mathbf{b} } \left( 1 + \frac{ e^{\mathbf{x_i}^\top \bm{\beta}} \; H_0(t_i | \bm{\lambda} ) }{\theta} \right) ^{-\theta}  \right] ^{ 1-\delta _{i} } 
\lambda_{k}^{a-1} e^{-b \lambda_{k}}.\label{posterior_lambdak_SMCFM}
\end{eqnarray}
The full conditional posterior distribution for $\theta$ is 
\begin{eqnarray}
\pi(\theta \mid \mathbf{b}, \bm{\beta}, \bm{\lambda}, \mathbf{D}) &\propto& 
\prod_{i=1}^{n}  
 \left( 1 + \frac{ e^{\mathbf{x_i}^\top \bm{\beta}} \; H_0(t_i | \bm{\lambda}) }{\theta} \right) ^{-(\theta+1) \delta _{i} } \notag \\
&\times& \left[ 1 + e^{ \mathbf{z}_{i}^\top \mathbf{b} } \left( 1 + \frac{ e^{\mathbf{x_i}^\top \bm{\beta}} \; H_0(t_i | \bm{\lambda} ) }{\theta} \right) ^{-\theta}  \right] ^{ 1-\delta _{i} } 
\theta^{c-1} e^{-d \theta}.  \label{posterior_theta_SMCFM}
\end{eqnarray}
The conditional posterior distributions of the remaining parameters, $\bm{\tau}, \bm{\tau}, \sigma^2, \sigma^{*2}, \eta^2$, and $\eta^{2}$, follow standard distributional forms, identical to those in the SMCM case (in Equations (\ref{posterior_tau}–\ref{posterior_eta_star}), and are accordingly sampled via Gibbs sampling.
To avoid redundancy, we do not provide the full MCMC algorithm for the SMCFM, as it largely follows the structure of the SMCM algorithm. The only modifications are the additional step for sampling the frailty parameter $\theta$ and minor adjustments in the conditional posterior distributions of $\mathbf{b}, \bm{\beta}$, and $\bm{\lambda}$. 
The full algorithm, implemented by combining Equations (\ref{posterior_bk_SMCFM}–\ref{posterior_theta_SMCFM}) with the MCMC procedure for the SMCM described in Section \ref{sec:SMC_MCMC}, is reported in Algorithm \ref{alg:MCMC_SMCFM} of the Appendix.

For comparison, the SMCFM is also implemented in \texttt{RStan} following the same approach used for the SMCM. The only modification is the inclusion of a prior for the frailty parameter $\theta$, as $\theta \sim \text{Gamma}(\theta_a, \theta_b)$, reflecting its role in the model.
As in the SMCM case, we implement the SMCFM in two variants: one following the hierarchical Bayesian structure and the other using standard normal priors for $\bm{b}$ and $\bm{\beta}$. 
These are denoted as \textit{HSMCFM(RStan)} and \textit{SMCFM(RStan)}, respectively. Both implementations are described via pseudo-codes in Algorithms \ref{alg:HSMCFM_Rstan} and \ref{alg:SMCFM_Rstan} of the Appendix.

\subsection{Bayesian Model Comparison Criteria} \label{sec:BayesianCriteria}

To evaluate and compare competing models, we employ the deviance information criterion (DIC) and the logarithm of the pseudo-marginal likelihood (LPML). These criteria are widely used for Bayesian cure models, as illustrated in recent studies such as \cite{ibrahim2012bayesian} and \cite{de2020bayesian}.

DIC, proposed by \cite{spiegelhalter2002bayesian}, balances model fit and complexity by incorporating both the deviance, which measures goodness of fit, and an effective number of parameters that account for model flexibility. 
The deviance is defined using the likelihood function of SMCM or SMCFM from (\ref{lik_SMCM}) or (\ref{lik_SMCFM})  as $\text{Dev}(\bm \nu) = -2 \log L(\bm \nu | \bm{D})$ where $\bm \nu = (\bm b, \bm \beta, \bm \lambda)$ or $ (\bm b, \bm \beta, \bm \lambda, \theta)$.
Let $\overline{\bm \nu}= E(\bm \nu | \bm D)$ and $\overline{\text{Dev}} = E\{ \text{Dev}(\bm \nu | \bm D) \}$ respectively denote the posterior means of $\bm \nu$ and $\text{Dev}(\bm \nu)$ with respect to the posterior distribution of all parameters.
The DIC measure is then defined as $\text{DIC} = \text{Dev}(\overline{\bm \nu}) + 2 p_D,$ where $p_D = \overline{\text{Dev}} - \text{Dev}(\overline{\bm \nu}) $ is the effective number of model parameters.
A lower DIC value indicates a better fit of the model to the data.
Note that the given form of DIC as a function of the effective number of model parameters is identical to AIC. However, unlike AIC, DIC uses the effective number of model parameters $p_D$ automatically derived from the posterior distribution.

LPML is a Bayesian goodness-of-fit measure based on the conditional predictive ordinate (CPO).
$\text{CPO}_i$ is the marginal posterior predictive density of the $i{\text{th}}$ sample given $\bm D_{(-i)}$, which indicates the observed data $\bm D$ with the $i{\text{th}}$ observation removed.
For the $i{\text{th}}$ observation, 
$$\text{CPO}_i =  \pi (t_i | \bm D_{(-i)} )= \int_\Theta L_i(t_i | \bm \nu ) \pi(\bm \nu | \bm D_{(-i)}) d\bm \nu = \bigg\{ \int_\Theta \frac{ \pi(\bm \nu | \bm D)}{L_i(t_i | \bm \nu )} d\bm \nu \bigg \}^{-1}$$ 
where $L_i(t_i | \bm \nu )$ is the likelihood contribution of the $i$th observation,  and $\pi(\bm \nu | \bm D_{(-i)} )$ denotes the joint posterior density of $\bm \nu $ based on the data $\bm D_{(-i)}$. 
According to \cite{chen2000monte}, a Monte Carlo approximation of $\text{CPO}_i$ is given by 
$$\widehat{\text{CPO}}_i = \left( \frac{1}{B} \sum_{j=1}^B \frac{1}{L( t_i| \bm \nu_j )}  \right)^{-1}$$
which is the harmonic mean of the likelihood values over the posterior draws $\{\bm \nu_j\}_{j=1}^B$ from the posterior distribution $\pi(\bm \nu | \bm D)$.
As suggested in \cite{ibrahim2001bayesian}, a useful summary statistic of the $\text{CPO}_i$ is the LPML defined as $\text{LPML} =  \sum_{i=1}^n \log \widehat{\text{CPO}}_i$. 
A larger value of LPML indicates a better fitting model.

In addition to DIC and LPML, we also consider the leave-one-out information criterion (LOOIC) implemented in the \texttt{loo} R package \cite{vehtari2017practical}. The LOOIC is based on Pareto-smoothed importance sampling leave-one-out cross-validation (PSIS-LOO-CV), which provides a computationally efficient approximation to exact leave-one-out cross-validation without requiring model refitting. Specifically, the method estimates the expected log predictive density for each observation by leaving it out of the fit, and the LOOIC is defined as $-2$ times this quantity, making it directly comparable to AIC and DIC.
As with AIC and DIC, lower values of the LOOIC indicate better model fit and predictive performance.
We use these model comparison criteria to determine the number of intervals $J$ for the baseline hazard function in both simulation and real data analyses.

\section{Simulation Study} \label{sec:simulation}

In this section, we present extensive simulation studies conducted to evaluate the performance of the proposed models.
For evaluating the results, we report the average posterior mean (only in the first scenario), the average posterior standard deviations (SDs), and the average Mean Absolute Error (MAE).
Except for the first scenario, each simulation setting involves at least $10$ unknown parameters. To summarize the results in a compact manner, we then display the SD and MAE values using bar plots, where each parameter is represented with a different color.
The MAE is employed as a measure of estimation accuracy, quantifying the absolute deviation between the true parameter values and their corresponding point estimates, i.e. the average posterior mean.
The MAE between the true parameter value $\theta_j  \in \bm{\theta}=(\theta_1,\cdots, \theta_p)$ and the corresponding point estimate $\hat{\theta}_j$ are evaluated for each $\theta_j$ as 
\begin{equation*}
        MAE(\theta_j, \hat{\theta}_j) = \frac{1}{N} \sum_{i=1}^{N} \mid \theta_j - \hat{\theta}_{ij} \mid , \; j=1,\cdots, p,
\end{equation*}
where $N$ is the total number of simulated datasets, and $\hat{\theta}_{ij}$ is the point estimate of $\theta_j$ obtained from the $i\text{th}$ simulated dataset.
All simulation scenarios are implemented in the R environment (version 4.1.2. \citeyear{R}), and the R code for implementing the proposed models algorithms' and for replicating the simulation studies is provided on GitHub at \url{https://github.com/fatihki/BayesSMCM/}.

The first simulation scenario is designed to mimic certain aspects of the studies developed by \cite{yin2005cure}, to facilitate a direct comparison of our respective models. 
In Scenario $1$, we assume an exponential distribution for the baseline survival function with $\lambda=1$, i.e. $J=1$. 
Two covariates, $\bm Z_1$ and $\bm Z_2,$ are independently generated respectively from a Bernoulli distribution with probability $0.5$, i.e. Bernoulli$(0.5),$ and from a standard normal distribution, i.e. $\mathcal{N}(0,1)$. The covariates $\bm X$ and $\bm Z$ are assumed to be the same.
The true regression coefficients are set as $(b_0, b_1, b_2) = (0.4, 0.5, 1)$ and $(\beta_1, \beta_2) = (1, 0.2)$. Following the specified parameter settings, the survival data is generated using SMCM. We set the number of observations to $n = 300, \; 500$ and $1000$.

In Scenario $2$, the piecewise exponential distribution is used as a baseline survival with $\bm \lambda = (0.2, 0.15, 0.3),$ hence $J=3$. The covariates in $\bm X$ and $\bm Z$ are not the same, and they are generated as: $\bm{X}_1 \sim \text{Bernoulli}(0.5)$,  $\bm{Z}_1 \sim \text{Bernoulli}(0.6),$ and $\bm{X}_2, \; \bm{X}_3, \; \bm{Z}_2, \; \bm{Z}_3   \sim  \mathcal{N}(0,1)$. The true regression coefficients are set as $(b_0, b_1, b_2, b_3) = (0.25, -1, 1.5, 0.5)$ and $(\beta_1, \beta_2, \beta_3) = (-1, 0.5, 2)$. The sample size is $n = 200, \; 400$ and $600$.

In Scenario $3$, similarly to Scenario $2$, it is assumed that $\bm \lambda = (0.15, 0.30, 0.50, 1),$ hence $J=4$. The covariates in $\bm X$ and $\bm Z$ are not the same, and they are generated as: $\bm{X}_1 \sim \text{Bernoulli}(0.5)$, $\bm{X}_2 \sim \text{Bernoulli}(0.25)$, $\bm{X}_3 \sim \text{Bernoulli}(0.65)$, $\bm{Z}_1 \sim \text{Bernoulli}(0.3)$, $\bm{Z}_2 \sim \text{Bernoulli(0.6)}$, and $\bm{X}_3, \; \bm{Z}_4  \sim  \mathcal{N}(0,1)$. 
The true regression coefficients are set as $(b_0, b_1, b_2, b_3) = ( -0.5, 1, 1.5, -2)$ and $(\beta_1, \beta_2, \beta_3, \beta_4) = (1.5, -0.30, 0.7, 1)$. 

In Scenario $4$, unlike the previous scenarios, the survival data is generated using a parametric MCM with a Weibull distribution, following a similar approach to \cite{zucknick2015nonidentical}. It is assumed that the covariates for the survival and latency parts are the same, thus $\bm X = \bm Z$, and generated as: $\bm{X}_1 \sim \text{Bernoulli}(0.8)$, $\bm{X}_2 \sim \text{Bernoulli}(0.3)$, $\bm{X}_3 \sim \text{Bernoulli}(0.4)$, $\bm{X}_4, \; \bm{X}_5  \sim  \mathcal{N}(0,1)$. 
The true regression coefficients are set as $(b_0, b_1, b_2, b_3, b_4) = ( 0.3, -1, 0.5, 1, 0.25)$ and $(\beta_1, \beta_2, \beta_3, \beta_4) = (-0.5, 1.5, 0.6, -0.8 )$. The sample size is $n = 200$ and $400$.
Since the data are generated from the parametric MCM, we consider different $J$ values for the baseline survival when applying our semiparametric models, with $J$ taking values in the set $\{1,2,3,4,5,7,10\}$.
Additionally, we compute model comparison criteria to evaluate the results for the different $J$ values, including classical measures such as AIC and BIC, as well as other criteria discussed in Section \ref{sec:BayesianCriteria}, DIC, LPML and LOOIC.

\subsection{Simulation results}\label{sec:simulation_results}

We here report the simulation results for each scenario specified above.
We repeat $500$ random simulations for all scenarios except Scenario $4$, for which we have $250$ replicates.
The average censoring (cure) rates in our simulated samples are approximately $37(35)\%, 75(66)\%, 47(40)\%, 65(50)\% $ for Scenario $1$ to $4$, respectively.
For each simulated data set, we run three chains with $15000$ iterations for all methods in the simulation. 
After discarding the first $2500$ iterations as burn-in, and applying a thinning of size $25$ to each chain, we retain $500$ iterations per chain, resulting in a total of $1500$ samples for posterior inference.
For monitoring the convergence of the posterior samples across all methods, we evaluate the estimated potential scale reduction factor (\cite{gelman1992inference}) for all unknown parameters. Based on our simulation results, this factor is generally around $1.00$ or below the commonly accepted threshold of $1.1$, indicating good convergence of the chains.

Table \ref{S1_table} presents the results of all proposed models for Scenario $1$, following the format of \cite{yin2005cure}. The table reports the average posterior means and SDs of the parameters, along with the uncured rate $\pi(\bm z)$. As expected, the posterior SDs decrease with increasing sample size.
When compared with the corresponding results from \cite{yin2005cure}, we observe that our models generally exhibit lower SD values and, in most cases, parameter estimates closer to the true values. In particular, the SMCM(RStan) and SMCFM(RStan) models provide the most accurate parameter estimates, except for the intercept term $b_0$, when compared to the other models.
Nevertheless, the other models also produce satisfactory estimates, with the exception of the coefficient $b_1$,
which tends to be underestimated in these models but overestimated in \cite{yin2005cure}.
Overall, it can be concluded that all models demonstrate comparable performance, and both model structures exhibit similar patterns with respect to the frailty component.

\begin{figure}[!h]
\centering
\includegraphics[width=0.95\textwidth]{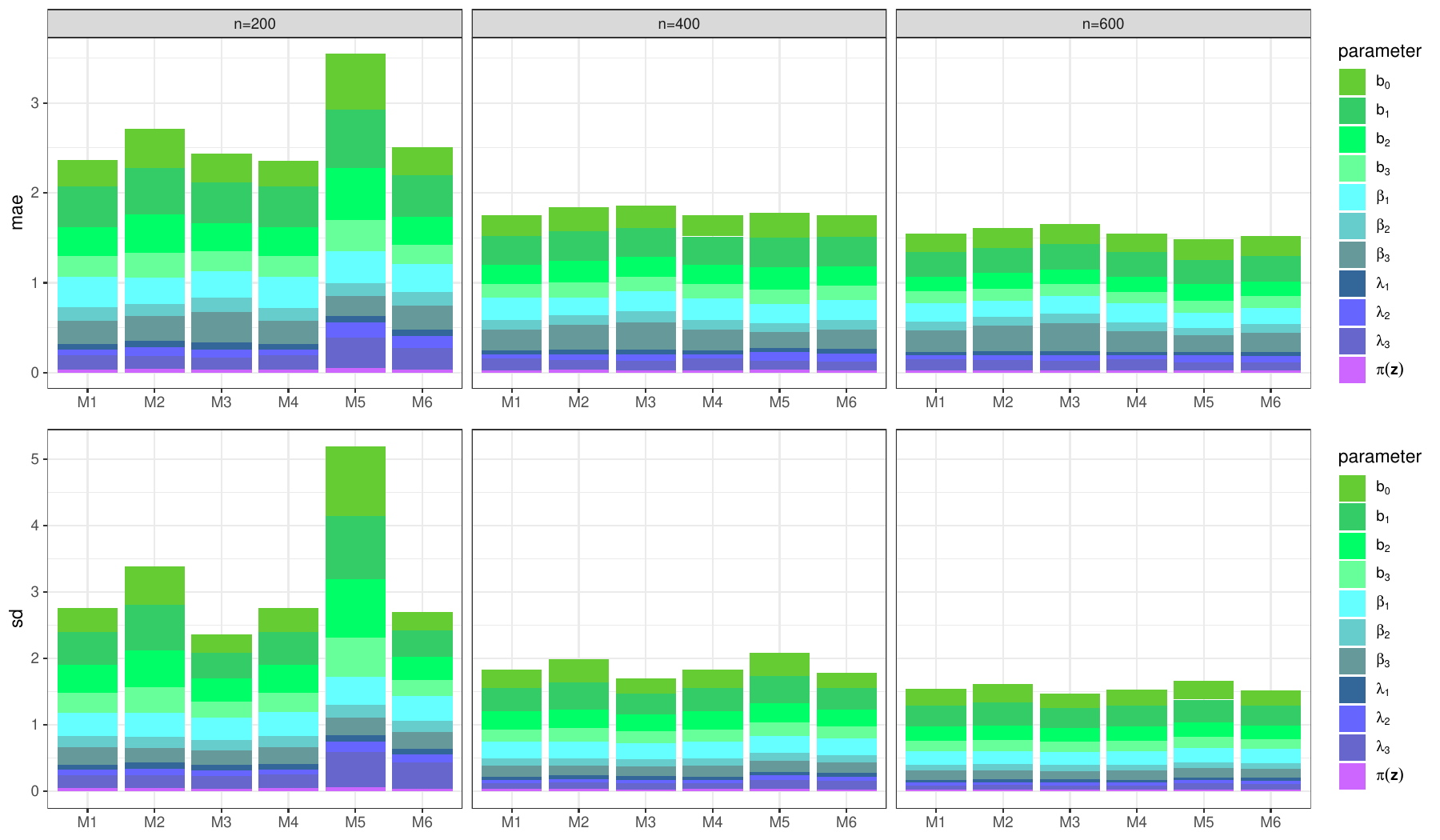}
\caption{Simulation results for Scenario $2$, Mean Absolute Error (MAE) and Standard Deviation (SD) for all parameter estimates.
Each bar represents the cumulative MAE (top panels) or SD (bottom panels) across all model parameters for methods M1-M6 and sample sizes $(n=200, 400, 600)$. Colors correspond to individual parameters. Methods M1–M6 correspond to SMCM(MCMC), SMCM(RStan), HSMCM(RStan), SMCFM(MCMC), SMCFM(RStan), and HSMCFM(RStan), respectively.}
\label{S2_mae_sd_plot}
\end{figure}

\begin{figure}[!h]
\centering
\includegraphics[width=0.95\textwidth]{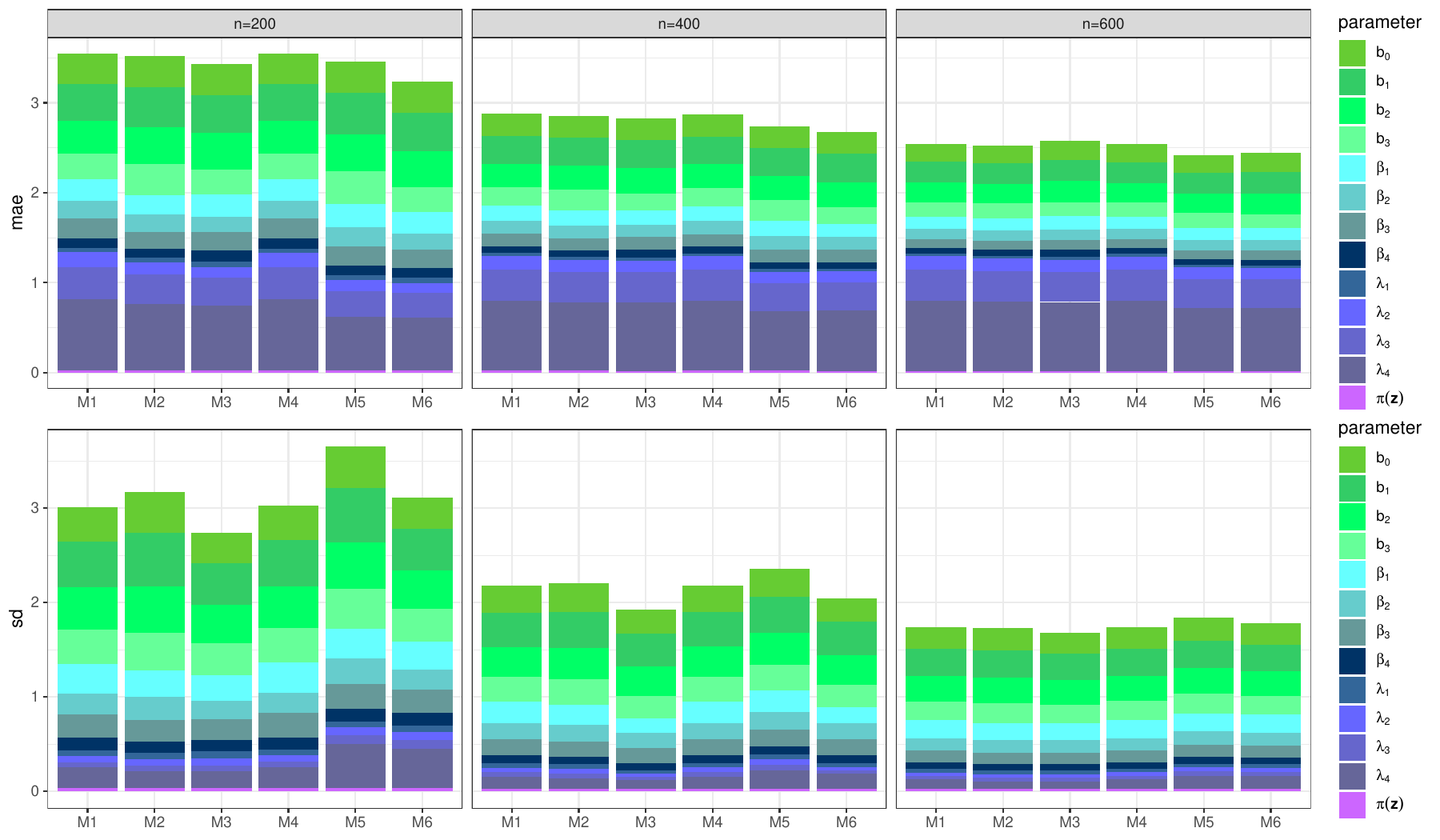}
\caption{Simulation results for Scenario $3$, MAE and SD plots for each parameter. Each bar represents the cumulative MAE and SD for methods M1–M6 and varying sample sizes.}
\label{S3_mae_sd_plot}
\end{figure}

Figures \ref{S2_mae_sd_plot} and \ref{S3_mae_sd_plot} compare the point estimation accuracy of each parameter in terms of MAE between the true values and the point estimates, as well as the corresponding SDs, for Scenarios $2$ and $3$, respectively.
We consider six models there, denoted as M1–M6, corresponding respectively to SMCM(MCMC), SMCM(RStan), HSMCM(RStan), SMCFM(MCMC), SMCFM(RStan), and HSMCFM(RStan).
These figures display the MAE (top panels) and posterior SD (bottom panels) of parameter estimates for Scenarios $2$ and $3$ across methods M1–M6 and sample sizes $n=200$, $400$, and $600$. 
As expected, all methods show improved estimation accuracy and stability as the sample size increases for both figures.

In Figure \ref{S2_mae_sd_plot}, methods M1, M3, M4, and M6 generally show comparable and good performance across all sample sizes, with consistently lower MAE and SD values compared to M2 and M5. While M1 exhibits slightly higher variability than M3, M4, and M6 in smaller samples, its overall estimation accuracy remains competitive.
The similar color composition across bars indicates that estimation errors are spread across parameters rather than dominated by a single coefficient. 
Nevertheless, parameters $b_0$ and $b_1$ contribute slightly more to the overall variability, consistent with the results in Table \ref{S1_table}.
In Figure \ref{S3_mae_sd_plot}, across all sample sizes, the overall MAE patterns indicate that models M1, M3, M4, and M6 perform comparably, with only slight differences observed between them.
In contrast, M2 and M5 exhibit similar MAE magnitudes to the better-performing models, but display noticeably higher variability in SD values. 
The largest contribution to total MAE appears to come from $\lambda_4$, possibly due to its relatively large true value and its location in the tail of the baseline hazard, where fewer events and higher censoring could increase estimation uncertainty.

Overall, inspection of Figures \ref{S2_mae_sd_plot} and \ref{S3_mae_sd_plot} shows that, interestingly, models M4–M6, which are based on SMCFMs, also demonstrate competitive and robust performance, even though the data were generated under the non-frailty SMCM framework. This suggests that the inclusion of a frailty term does not substantially deteriorate parameter estimation and may even enhance robustness against unobserved heterogeneity.
Across all methods, the main contributions to total MAE and SD generally arise from the regression coefficients of $\bm{b}$, while the uncured rate $\pi(\bm{z})$ is estimated with relatively smaller errors.
Overall, the results indicate that all models perform comparably at moderate to large sample sizes, with the non-frailty models (M1–M3) retaining a slight efficiency advantage, whereas the frailty-based models (M4–M6) exhibit good adaptability and stability even under model misspecification.
Moreover, M1, M3, M4, and M6 achieve reliable estimation accuracy even at moderate sample sizes, whereas M2 and M5 show higher variability, particularly when the sample size is small.

\begin{figure}[!h]
\centering
\includegraphics[width=0.95\textwidth]{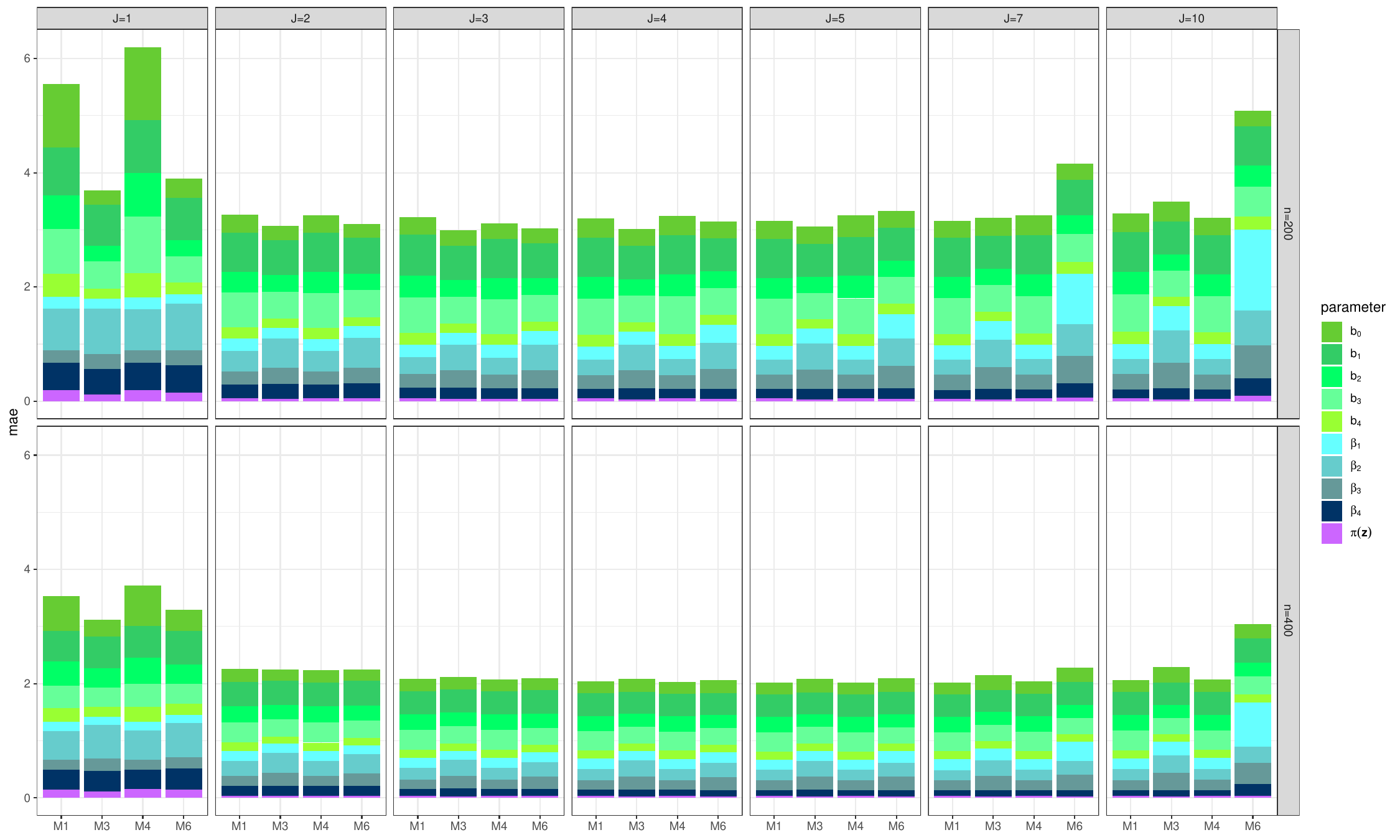}
\caption{Simulation results for Scenario $4$, MAE plots for each parameter. Each bar shows the cumulative MAE for methods M1, M3, M4, and M6 across different sample sizes, with $J \in \{1,2,3,4,5,7,10\}$.}
\label{S4_mae_plot}
\end{figure}

\begin{figure}[!h]
\centering
\includegraphics[width=0.95\textwidth]{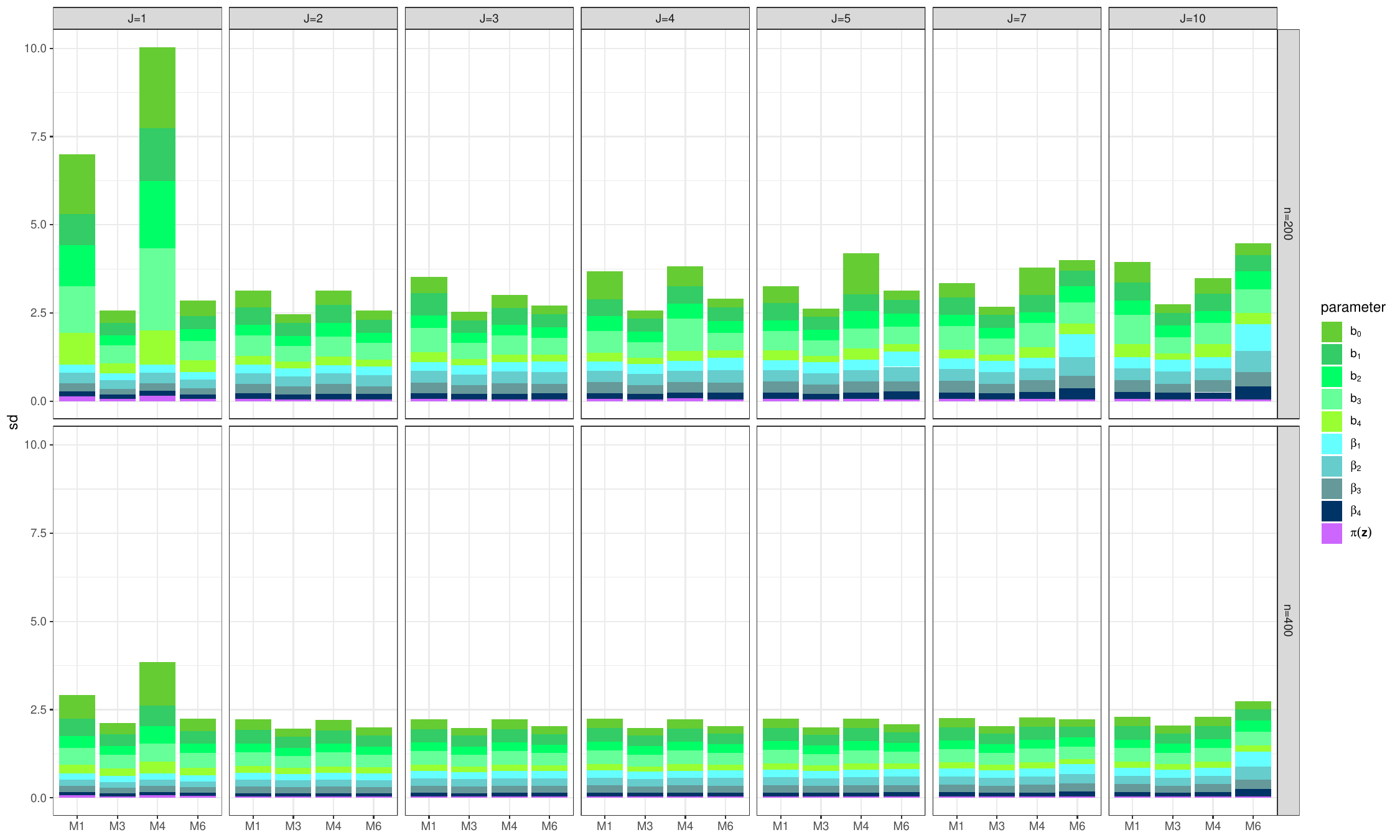}
\caption{Simulation results for Scenario $4$, SD plots for each parameter. 
Each bar shows the cumulative SD for methods M1, M3, M4, and M6 across different sample sizes, with $J \in \{1,2,3,4,5,7,10\}$.}
\label{S4_sd_plot}
\end{figure}

\begin{figure}[!h]
\centering
\includegraphics[width=1\textwidth]{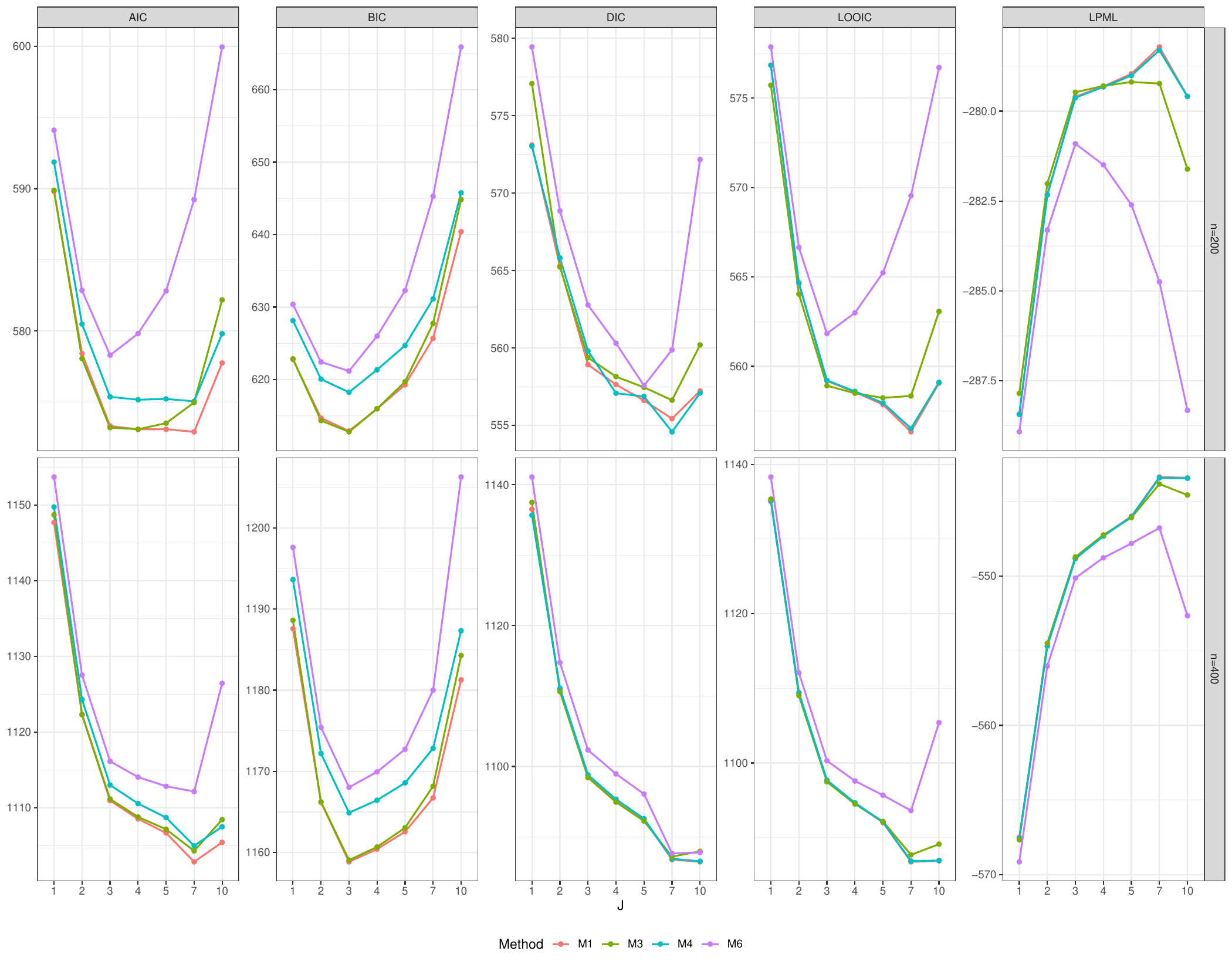}
\caption{Simulation results for Scenario $4$. Plots of Bayesian model selection criteria, illustrating the performance of models M1, M3, M4, and M6 across different sample sizes, with $J \in \{1,2,3,4,5,7,10\}$.}
\label{S4_Bayes_criteria_plot}
\end{figure}

Figures \ref{S4_mae_plot}, \ref{S4_sd_plot}, and \ref{S4_Bayes_criteria_plot} present MAE and posterior SD plots similar to the previous figures, along with Bayesian model selection criteria for Scenario $4$, across methods M1, M3, M4, and M6, and sample sizes $n = 200, 400$ with $J \in \{1,2,3,4,5,7,10\}$. Since models M2 and M5 exhibit larger errors for small samples and show little difference from other models for larger samples, and since they are based on \texttt{RStan} as models M3 and M6, M2 and M5 are not evaluated under this simulation scenario.

Figures \ref{S4_mae_plot} and \ref{S4_sd_plot} shows that the cumulative MAE and SD values generally decrease as the number of time intervals $J$ increases from $1$ to $\{2, 3, 4, 5, 7\}$, indicating improved estimation accuracy and stability as the piecewise baseline hazard becomes more flexible. However, the overall MAE and SD patterns remain quite similar among models for $J=2,3,4,5,7$, except for M6, which shows slightly higher cumulative errors. When $J$ increases further to $10$, both MAE and SD values rise markedly, particularly for M3 and M6, indicating mild overfitting and reduced estimation efficiency when too many time intervals are used to approximate the baseline hazard.
Considering model structures, M1 and M3 (based on SMCM) and M4 and M6 (based on SMCFM) display very similar MAE and SD performance for $J=2,3,4,5$ and under the larger sample size ($n=400$), indicating that the inclusion of a frailty term does not substantially affect estimation quality in well-informed settings. In contrast, for smaller samples ($n=200$), the SMCFM-based models show slightly larger MAE values and more noticeable variability in SD, reflecting the increased estimation uncertainty associated with fitting additional frailty parameters under limited data.

Figure \ref{S4_Bayes_criteria_plot} further supports these findings through the Bayesian model selection criteria (AIC, BIC, DIC, LOOIC, and LPML), all of which exhibit a distinct U-shaped pattern with respect to $J$ in almost all cases. 
The best overall model performance is generally achieved at $J=3,7$ for AIC, at $J=3$ for BIC, and at 
$J=7$ for DIC, LOOIC, and LPML across all applied models. 
Because AIC and BIC tend to penalize model complexity more strongly, they may favor overly simplistic specifications in SMCMs. Therefore, greater emphasis is placed on the DIC, LOOIC, and LPML criteria when assessing model adequacy. Consistently with the estimation results, M1 and M3 generally yield the most favorable criterion values across all $J$ and sample sizes, confirming their overall superiority. 
M4 also produces criterion curves highly comparable to those of M1 and M3, whereas M6 is clearly distinguished from the others, showing less favorable results.
The data for this scenario were generated under a parametric MCM with a Weibull baseline distribution. Accordingly, under model misspecification, the SMCM-based models (M1 and M3) generally exhibit better performance than the SMCFM-based model (M6). However, M4, which employs its own MCMC estimation approach, achieves performance comparable to that of M1 and M3.
Collectively, these results indicate that moderate hazard flexibility and adequate sample size produce the most accurate and stable estimates, with M1 and M3 offering the best balance between bias, variance, and model complexity.

Overall, inspection of Figures \ref{S4_mae_plot}, \ref{S4_sd_plot}, and \ref{S4_Bayes_criteria_plot} reveals that, as expected, increasing the sample size consistently reduces estimation variability across all methods. The models with $J=7$ provide the best balance between flexibility and parsimony, yielding the most accurate and stable estimates across both SMCM- and SMCFM-based approaches, with the exception of model M6.
\begin{table}[!h]
	\centering
 \renewcommand{\arraystretch}{1.40}
 \caption{Simulation results for Scenario $1$. Posterior mean and standard deviation (SD) for all parameters $\mathbf{b},\: \bm{\beta}, \; \lambda$ and $\pi(\mathbf{z})$ obtained with all proposed methods (M1-M6) and with $n=300,500,1000.$ The true values of $\pi(\bm z)$ are respectively 0.65072, 0.6512, and 0.65333 for the three values of $n$.}
 \label{S1_table}
 \resizebox{1\textwidth}{!}{
	\begin{tabular}{ccccccccc|ccccccc} \hline
     $n$ & Estimate & $b_0=0.4$ & $b_1=0.5$ & $b_2=0.1$ & $\beta_1=1$ & $\beta_2=0.2$ & $\lambda=1$ & $\pi(\bm z)$ & $b_0=0.4$ & $b_1=0.5$ & $b_2=0.1$ & $\beta_1=1$ & $\beta_2=0.2$ & $\lambda=1$ & $\pi(\bm z)$ \\ \hline
     \multicolumn{2}{c}{} & \multicolumn{7}{c@{\quad}}{M1: SMCM(MCMC)} & \multicolumn{7}{c@{\quad}}{M4: SMCFM(MCMC)}  \\
     \cmidrule(r){3-9} \cmidrule(r){10-16}
     300 & Mean & 0.40721 & 0.41809 & 0.08780 & 0.88866 & 0.18222 & 1.06389 & 0.64469 & 0.40813 & 0.41799 & 0.08773 & 0.88668 & 0.18171 & 1.07644 & 0.64487  \\ 
          & SD & 0.17135 & 0.21789 & 0.11057 & 0.18463 & 0.07349 & 0.19025 & 0.03090  & 0.17155 & 0.21859 & 0.11092 & 0.18481 & 0.07331 & 0.19371 & 0.03080 \\ 
    500 & Mean & 0.41221 & 0.44031 & 0.08968 & 0.91702 & 0.18430 & 1.04066 & 0.64853 & 0.41262 & 0.44006 & 0.08956 & 0.91620 & 0.18417 & 1.04741 & 0.64859 \\  
         & SD & 0.14250 & 0.19054 & 0.08886 & 0.15495 & 0.05904 & 0.16400 & 0.02413 & 0.14242 & 0.19091 & 0.08870 & 0.15493 & 0.05891 & 0.16528 & 0.02407  \\ 
    1000 & Mean & 0.40900 & 0.46298 & 0.09167 & 0.93954 & 0.18595 & 1.01291 & 0.65181 & 0.40894 & 0.46306 & 0.09164 & 0.93939 & 0.18598 & 1.01591 & 0.65181 \\  
         & SD & 0.10171 & 0.14127 & 0.06743 & 0.10417 & 0.03935 & 0.10571 & 0.01694 & 0.10190 & 0.14176 & 0.06754 & 0.10380 & 0.03926 & 0.10615 & 0.01694 \\ \hline 
         \multicolumn{2}{c}{} & \multicolumn{7}{c@{\quad}}{M2: SMCM(RStan)}  & \multicolumn{7}{c@{\quad}}{M5: SMCFM (RStan)}  \\
     \cmidrule(r){3-9} \cmidrule(r){10-16}
     300 & Mean & 0.42557 & 0.49988 & 0.10853 & 0.91934 & 0.19251 & 1.05060 & 0.65650 & 0.42879 & 0.49941 & 0.10835 & 0.91477 & 0.19155 & 1.10499 & 0.65714 \\  
         & SD & 0.19188 & 0.25313 & 0.13543 & 0.18008 & 0.07438 & 0.18908 & 0.02887 & 0.19273 & 0.25455 & 0.13565 & 0.17947 & 0.07434 & 0.20081 & 0.02899 \\
     500 & Mean & 0.41963 & 0.49432 & 0.10538 & 0.93295 & 0.19021 & 1.03439 & 0.65534 & 0.42210 & 0.49385 & 0.10513 & 0.92937 & 0.18974 & 1.07849 & 0.65583 \\  
         & SD & 0.15074 & 0.20346 & 0.10191 & 0.15369 & 0.05927 & 0.16444 & 0.02335 & 0.15127 & 0.20437 & 0.10220 & 0.15354 & 0.05899 & 0.17158 & 0.02337 \\  
    1000 & Mean & 0.41019 & 0.49353 & 0.10195 & 0.94742 & 0.18880 & 1.00982 & 0.65510 & 0.41211 & 0.49286 & 0.10178 & 0.94518 & 0.18822 & 1.04284 & 0.65546 \\ 
         & SD & 0.10296 & 0.14240 & 0.07266 & 0.10347 & 0.03924 & 0.10582 & 0.01669 & 0.10304 & 0.14247 & 0.07267 & 0.10325 & 0.03899 & 0.10934 & 0.01665 \\ \hline
                \multicolumn{2}{c}{} & \multicolumn{7}{c@{\quad}}{M3: HSMCM(RStan)}  & \multicolumn{7}{c@{\quad}}{M6: HSMCFM (RStan)} \\
     \cmidrule(r){3-9} \cmidrule(r){10-16}
     300 & Mean & 0.40441 & 0.41100 & 0.08521 & 0.85469 & 0.17702 & 1.09028 & 0.64349 & 0.40731 & 0.41141 & 0.08529 & 0.84240 & 0.17438 & 1.15291 & 0.64419 \\ 
         & SD & 0.16682 & 0.20994 & 0.10638 & 0.17862 & 0.07291 & 0.18877 & 0.02915 & 0.16751 & 0.21092 & 0.10654 & 0.17865 & 0.07235 & 0.20159 & 0.02917 \\
     500 & Mean & 0.40977 & 0.43472 & 0.08768 & 0.89621 & 0.18109 & 1.05680 & 0.64748 & 0.41181 & 0.43452 & 0.08746 & 0.88875 & 0.17963 & 1.10425 & 0.64792 \\ 
         & SD & 0.14121 & 0.18668 & 0.08633 & 0.15191 & 0.05896 & 0.16261 & 0.02372 & 0.14165 & 0.18741 & 0.08663 & 0.15125 & 0.05860 & 0.16960 & 0.02375 \\ 
    1000 & Mean & 0.40770 & 0.45939 & 0.09024 & 0.92902 & 0.18431 & 1.02089 & 0.65117 & 0.40930 & 0.45932 & 0.09022 & 0.92562 & 0.18341 & 1.05518 & 0.65152 \\  
         &  SD & 0.10166 & 0.14021 & 0.06647 & 0.10325 & 0.03939 & 0.10546 & 0.01688 & 0.10145 & 0.14067 & 0.06652 & 0.10301 & 0.03913 & 0.10916 & 0.01683 \\  \hline
     	\end{tabular}
        }
\end{table}

\clearpage
\section{Data Analysis} \label{sec:data_analysis}

In this section, we analyze two benchmark datasets to illustrate the proposed mixture cure model approaches. Specifically, we consider the E1690 melanoma clinical trial dataset and the colon cancer clinical trial dataset, which serve as representative examples for evaluating model performance in practice.
The detailed analyses of each dataset are presented in the following subsections.

\subsection{ECOG E1690 clinical trial data}

The ECOG E1690 trial evaluated the efficacy of Interferon Alfa-2b (IFN) as adjuvant therapy following surgical resection of deep primary or regionally metastatic melanoma, with a comprehensive account provided by \cite{kirkwood2000high}.
This dataset has been widely used to illustrate mixture cure models and related approaches, including those by \cite{yin2005cure}, \cite{ibrahim2012bayesian}, \cite{psioda2018bayesian}, and, more recently, \cite{de2020bayesian}.
For direct comparability with the results of \cite{yin2005cure}, we adopt the same data structure as in their analysis. The E1690 dataset was retrieved from the GitHub repository provided by \cite{psioda2018bayesian}.

\begin{figure}[!h]
\centering
\includegraphics[width=1\textwidth]{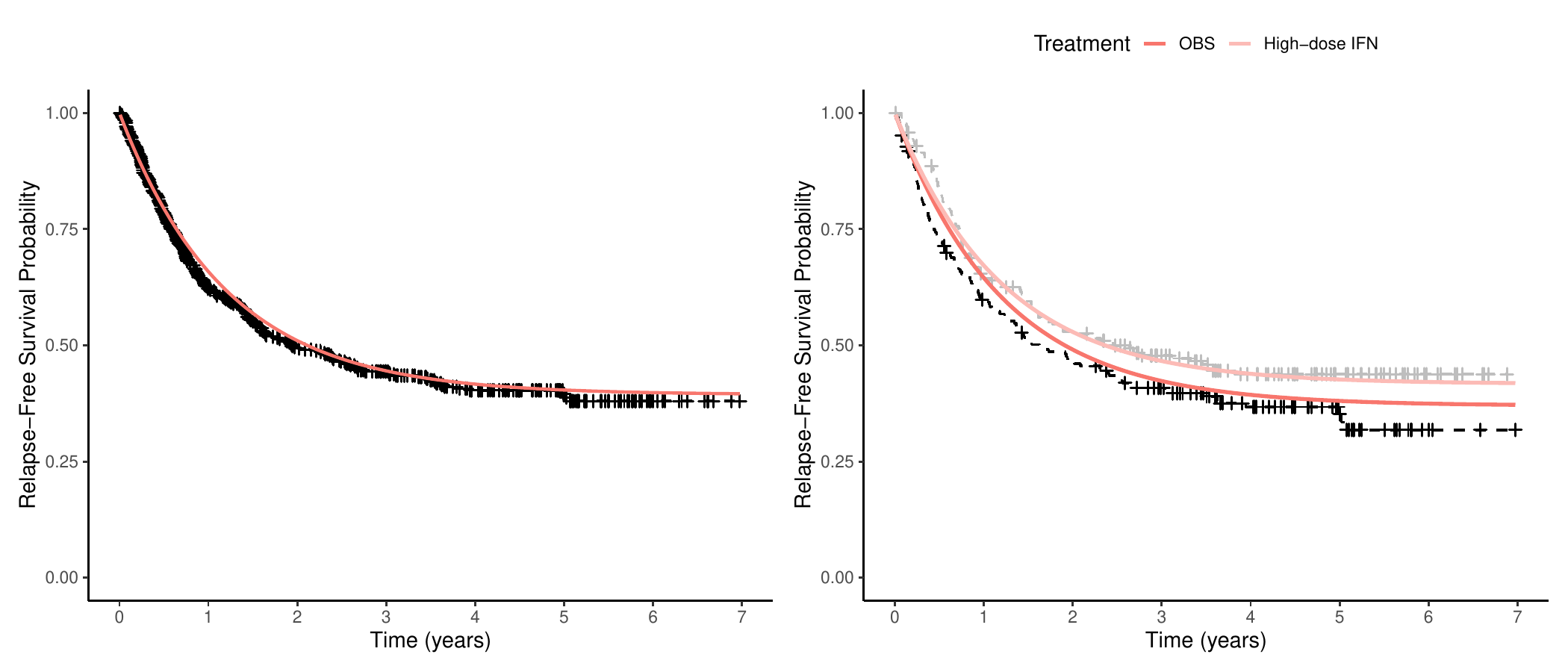}
\caption{Results of the analysis of the E1690 dataset. Kaplan–Meier estimates of survival (black) and posterior survival functions (red) obtained from the selected SMCM(MCMC) ($J=1$) best model. The left panel shows overall survival, while the right panel presents survival curves stratified by treatment.}
\label{Data1_KM_Model_plots}
\end{figure}

The outcome of interest is relapse-free survival (RFS), defined as the time from randomization until tumor progression or death, whichever occurs first \cite{ibrahim2012bayesian}.
After excluding $10$ patients with zero RFS, the dataset includes $n=417$ patients, with approximately $42\%$ censoring across the two treatment arms.
The dataset in this application comprises: $t:$ observed RFS in years (mean $2.31$ years); $\delta:$ censoring indicator ($177$ right-censored observations); $x_1:$ treatment group ($212$ patients in the high-dose IFN arm = 1, $205$ in the observation arm = 0); $x_2:$ age (continuous, range $19.13$–$78.05$, mean $48.05$ years); and $x_3:$ sex ($261$ male = 0, $156$ female = 1).
For the incidence part of the model, we use the same covariates in addition to the intercept, i.e., $\bm{z} = [\bm{1}, \bm{x}]$.
Among patients who did not experience relapse, both the median and mean follow-up times exceeded four years. Figure \ref{Data1_KM_Model_plots} displays Kaplan–Meier (KM) estimates of the survival curves overall and stratified by treatment group (high-dose IFN vs. observation). A clear plateau at approximately four years suggests the presence of long-term survivors, motivating the use of a cure rate model for this dataset.

We apply four proposed mixture cure model approaches -- SMCM (MCMC), SMCFM (MCMC), HSMCM (RStan), and HSMCFM (RStan) -- to the E1690 dataset. As in \cite{yin2005cure}, we set $J \in \{1,2,3,4\}$ for the semiparametric structure. For each method, we run five MCMC chains of $60000$ iterations, discarding the first $10000$ as burn-in and applying a thinning by $50$, resulting in $5000$ posterior samples in total.
Model comparisons based on BIC, DIC, and LPML are reported in Table \ref{tab:Bayes_comparisons_E1690data}, while posterior means, standard deviations, and HPD intervals for all parameters are given in Table \ref{tab:Bayes_estimates_E1690data}.

Table \ref{tab:Bayes_comparisons_E1690data} shows that, according to all three criteria, the best-fitting models correspond to the case $J=1$ across all approaches, consistent with the results of \cite{yin2005cure}, with SMCM(MCMC) demonstrating the best overall performance.
Figure \ref{Data1_KM_Model_plots} also shows the posterior mean survival curves of the best model for both overall survival and survival stratified by treatment. 
The posterior survival curves (in red) closely align with the empirical Kaplan–Meier estimates (in black), both overall and stratified by treatment. This agreement indicates that the proposed SMCM(MCMC) model captures the survival dynamics and treatment effects in the E1690 dataset well, providing a satisfactory fit for inference and prediction.
Table \ref{tab:Bayes_estimates_E1690data} indicates that compared with \cite{yin2005cure}, our results generally exhibit smaller posterior standard deviations, with narrower credible intervals, indicating improved estimation precision. In contrast to their findings, however, age does not emerge as an important covariate in both parts of the model for long-term survival in our analysis.

\begin{table}[!h]
\centering
 \renewcommand{\arraystretch}{1.30}
 \caption{Results of the analysis of the E1690 dataset. Bayesian model comparison criteria for all fitted models.} 
  \resizebox{1\textwidth}{!}{
\begin{tabular}{ccccccc}
\toprule 
$J$ & Criteria & \shortstack{ Results from \\ \cite{yin2005cure} }&SMCM(MCMC) & SMCFM(MCMC) & HSMCM(RStan) & HSMCFM(RStan) \\ \hline 
1 & BIC & -& \textbf{1081.0784} & 1087.1704 & 1081.9513 & 1087.1718\\
  & DIC & -& \textbf{1037.4985} & 1037.6609 & 1039.5167& 1039.3066\\ 
  & LPML& -522.08 & \textbf{-518.7946} & -518.8818 & -519.8967 & -519.4694 \\ \hline
2 & BIC & - &1087.7268 & 1093.7487&  1088.3520 & 1093.9454\\
  & DIC & - &1039.2362 & 1039.2764&  1040.7602 & 1040.0788\\
  & LPML& -523.05 &-519.6820 & -519.7009& -520.5138 & -520.3819 \\\hline  
3 & BIC & -&1094.6380 &  1100.6841 & 1095.2151 & 1099.9361\\
  & DIC & -&1041.1833 & 1041.2393 & 1042.6774 & 1037.5168 \\ 
  & LPML& -522.60&-520.6507 & -520.6787 & -521.4950 & -520.9072\\\hline   
4 & BIC &  -&1099.4886 &  1105.1884 & 1099.1591 & 1105.4917 \\
  & DIC &  -&1040.0720 &  1040.7568 &  1041.5934 & 1038.3784\\ 
  & LPML& -524.18 &-520.8744  & -520.5779 & -520.8894 &  521.0682\\\hline    
\end{tabular}
}
\label{tab:Bayes_comparisons_E1690data}
\end{table}

\begin{sidewaystable}[!h]
\centering
 \renewcommand{\arraystretch}{1.20}
\caption{Results of the analysis of the E1690 dataset. Posterior estimates of all parameters for all fitted models (Est: posterior mean; SD: posterior standard deviation; HPD: $95\%$ High-Posterior Density credible interval)}
\resizebox{0.95\textheight}{!}{
\begin{tabular}{ccccc ccc ccc ccc}
\toprule
 & & 
\multicolumn{3}{c}{SMCM(MCMC)} &
\multicolumn{3}{c}{SMCFM(MCMC)} &
\multicolumn{3}{c}{HSMCM(RStan)} &
\multicolumn{3}{c}{HSMCFM(RStan)} \\
\cmidrule(lr){3-5}
\cmidrule(lr){6-8}
\cmidrule(lr){9-11}
\cmidrule(lr){12-14}
$J$ & parameter & Est & SD & HPD & Est & SD & HPD & Est & SD & HPD & Est & SD & HPD \\
\midrule
1 &  $b_0$ & 0.5660 & 0.1848 & (0.2100,0.9280) & 0.5683 & 0.1878 & (0.2080,0.9363) &
0.5584 & 0.1743 & (0.2320, 0.9084) & 0.6896 & 0.2398 & (0.2735, 1.1187)\\
&   $b_1$ & -0.1510 & 0.1938 & (-0.5609,0.1837) & -0.1482 & 0.1903 & (-0.5428,0.1939) &-0.1344 & 0.1836 & (-0.4996,0.2073) &  -0.1366 & 0.1994 & (-0.5537,0.2216)\\
&   $b_2$ & 0.1914 & 0.1087 & (-0.0169,0.3996) & 0.1902 & 0.1112 & (-0.0205,0.4023) &0.1998 & 0.1102 & (-0.0160,0.4110)&  0.2240 & 0.1294 & (-0.0217,0.4661)\\
&   $b_3$ & -0.1258 & 0.1836 & (-0.5141,0.2129)& -0.1240 & 0.1851 & (-0.5178,0.2051) &-0.1127 & 0.1807 & (-0.5102,0.2005)& -0.1304 & 0.1978 & (-0.5378,0.2297)\\
&   $\beta_1$ & -0.0012 & 0.0262 & (-0.0504,0.0354)  & -0.0018 & 0.0304 &(-0.0699,0.0457) &-0.0261 & 0.1058 & (-0.2541,0.1841)& -0.0661 & 0.1325 &(-0.3674,0.1594)\\
&   $\beta_2$ & -0.0026 & 0.0231 & (-0.0490,0.0355) & -0.0029 & 0.0240 & (-0.0550,0.0392) &-0.0310 & 0.0737 & (-0.1874,0.1066)& -0.0292 & 0.0819 & (-0.2047,0.1271)\\
&   $\beta_3$ & -0.0008 & 0.0270 & (-0.0406,0.0495) & -0.0021 & 0.0306 & (-0.0620,0.0531)&-0.0206 & 0.1083 & (-0.2537,0.1927)& -0.0190 & 0.1251 & (-0.2919,0.2300)\\
&   $\lambda_1$ & 0.8320 & 0.0684 & (0.7035,0.9693) & 0.8343 & 0.0691 & (0.6953,0.9659) & 0.8501 & 0.0857 & (0.6831,1.0186)& 0.9254 & 0.1211 & (0.6944,1.1616)\\  
&  $\pi(\bm z)$ & 0.6074 & 0.0272 & (0.5506,0.6583) & 0.6085 & 0.0280 & (0.5518,0.6604) & 0.6088 & 0.0269 & (0.5533,0.6584)& 0.6370 & 0.0383 & (0.5635,0.7056) \\  \hline
2 & $b_0$ & 0.5898 & 0.1890 & (0.2343,0.9650)& 0.5964 & 0.1945 & (0.2503,1.0140) &
 0.5786 & 0.1771 & (0.2480,0.9339) & 0.7232 & 0.2706 & (0.2619,1.2247)\\
&   $b_1$ & -0.1511 & 0.1936 & (-0.5388,0.2179) & -0.1550 & 0.1938 & (-0.5606,0.1904) &-0.1330 & 0.1853 & (-0.5193,0.1937)&  -0.1386 & 0.2046 & (-0.5788,0.2226)\\
&   $b_2$ & 0.1961 & 0.1124 & (-0.0179,0.4180) & 0.1962 & 0.1133 & (-0.0155,0.4178) & 0.2034 & 0.1118 & (-0.0124,0.4188)&  0.2322 & 0.1331 & (0.0006,0.5031)\\
&   $b_3$ & -0.1286 & 0.1857 & (-0.5198,0.2080) & -0.1287 & 0.1903 &(-0.5087,0.2356) &  -0.1149 & 0.1827 & (-0.5222,0.1930)&  -0.1381 & 0.2060 & (-0.5588,0.2514)\\
&   $\beta_1$ & -0.0019 & 0.0330 & (-0.0781,0.0518) & -0.0026 & 0.0349 & (-0.0899,0.0591)&-0.0294 & 0.1064 & (-0.2664,0.1802)& -0.0776 & 0.1399 & (-0.3969,0.1601) \\
&   $\beta_2$ & -0.0034 & 0.0273 & (-0.0681,0.0476) & -0.0036 & 0.0276 & (-0.0725,0.0507) &-0.0298 & 0.0731 & (-0.1863,0.1055) & -0.0338 & 0.0848 & (-0.2256,0.1202)\\
&   $\beta_3$ & -0.0007 & 0.0349 & (-0.0658,0.0615) & -0.0012 & 0.0379 & (-0.0634,0.0777) &-0.0195 & 0.1108 & (-0.2649,0.2011) &  -0.0144 & 0.1297 & (-0.2990,0.2480) \\
&   $\lambda_1$ & 0.8525 & 0.0843 & (0.6858,1.0148) &  0.8523 & 0.0840 & (0.7052,1.0268)&0.8776 & 0.0996 & (0.6887,1.0726)& 0.9256 & 0.1259 & (0.6918,1.1759) \\
&  $\lambda_2$ & 0.7675 & 0.1021 & (0.5617,0.9659) & 0.7677 & 0.1035 &(0.5766,0.9799)&0.7932 & 0.1112 & (0.5859,1.0170)&  1.0701 & 0.2772 & (0.6010,1.6063)\\  
&  $\pi(\bm z)$ & 0.6127 & 0.0285 & (0.5557,0.6680) & 0.6138 & 0.0288 & (0.5569,0.6692) & 0.6134 & 0.0277 & (0.5562,0.6659)&   0.6436 & 0.0433 & (0.5636,0.7283) \\  \hline
3  & $b_0$ &0.6238 & 0.2068 & (0.2184, 1.0059) & 0.6236 & 0.2096 & (0.2489,1.0344)&0.6012 & 0.1846 & (0.2587,0.9766) & 0.8127 & 0.3737 & (0.2705,1.5157) \\
&   $b_1$ & -0.1568 & 0.1989 & (-0.5687,0.1974) & -0.1577 & 0.1963 & (-0.5758,0.1821)&-0.1379 & 0.1914 & (-0.5417,0.2028)& -0.1443 & 0.2213 & (-0.6107,0.2555) \\
&   $b_2$ & 0.2034 & 0.1142 & (-0.0162,0.4224) & 0.2046 & 0.1165 & (-0.0180,0.4237) &0.2107 & 0.1162 & (-0.0136,0.4281)& 0.2530 & 0.1647 & (-0.0278,0.5332)\\
&   $b_3$ & -0.1345 & 0.1936 & (-0.5148,0.2425) & -0.1352 & 0.1950 & (-0.5242,0.2328) & -0.1201 & 0.1842 & (-0.4835,0.2356)& -0.1476 & 0.2285 & (-0.6078,0.2640)\\
&   $\beta_1$ & -0.0016 & 0.0317 & (-0.0675,0.0570) & -0.0008 & 0.0300 & (-0.0626,0.0543) &-0.0350 & 0.1112 & (-0.2763,0.1742)& -0.1021 & 0.1612 & (-0.4559,0.1727) \\
&   $\beta_2$ & -0.0035 & 0.0260 & (-0.0616,0.0453) & -0.0031 & 0.0264& (-0.0553,0.0377) & -0.0293 & 0.0746 & (-0.1959,0.1100) & -0.0329 & 0.0883 & (-0.2157,0.1346)\\
&   $\beta_3$ & -0.0006 & 0.0325 & (-0.0560,0.0678) & -0.0001 & 0.0311& (-0.0719,0.0517)&-0.0184 & 0.1112 & (-0.2666,0.1943)& -0.0098 & 0.1357 & (-0.2928,0.2714)\\
&   $\lambda_1$ & 0.8218 & 0.0947 & (0.6429,1.0130) &  0.8203 & 0.0941 & (0.6455,1.0184)& 0.8499 & 0.1090 & (0.6330,1.0589)&  0.8962 & 0.1418 & (0.6331,1.1726)\\
&   $\lambda_2$ & 0.8308 & 0.1042 & (0.6337,1.0362) & 0.8354 & 0.1055 & (0.6443,1.0576) &0.8654 & 0.1156 & (0.6422,1.0927)& 1.1911 & 0.3176 & (0.7027,1.8270)\\
&   $\lambda_3$ & 0.7076 & 0.1395 & (0.4575,0.9964) & 0.7095 & 0.1420 & (0.4383,0.9853) &0.7436 & 0.1440 & (0.4750,1.0307)& 1.4567 & 0.7355 & (0.4955,2.8441) \\   
&   $\pi(\bm z)$ & 0.6193 & 0.0311 & (0.5563,0.6777) & 0.6191 & 0.0317 & (0.5538,0.6772) & 0.6176 & 0.0299 & (0.5621,0.6779) & 0.6617 & 0.0552 & (0.5614,0.7682)\\  \hline
4 &  $b_0$ & 0.9059 & 1.4901 & (0.1791,1.4513) & 0.8112 & 0.6869 & (0.2022,1.3881)&
0.6444 & 0.2159 & (0.2592,1.0264) & 0.8393 & 0.4014 & (0.2373,1.5501) \\
&   $b_1$ &  -0.1901 & 0.4636 & (-0.6837,0.2388) & -0.1821 & 0.2915 & (-0.6636,0.2221)&-0.1367 & 0.1964 & (-0.5421,0.2306)& -0.1468 & 0.2300 & (-0.6247,0.2697)\\
&   $b_2$ & 0.2373 & 0.2934 & (-0.0618,0.5122) & 0.2435 & 0.2200 & (-0.0417,0.5047)&0.2208 & 0.1313 & (-0.0249,0.4700)& 0.2518 & 0.1690 & (-0.0471,0.5566)\\
&   $b_3$ &-0.1877 & 0.6438 & (-0.6576,0.2649) & -0.1771 & 0.3265 & (-0.6330,0.2536)& -0.1237 & 0.1959 & (-0.5178,0.2492)& -0.1551 & 0.2357 & (-0.6273,0.2524)\\
&   $\beta_1$ & -0.0027 & 0.0372 & (-0.0859,0.0651) & -0.0029 & 0.0355 & (-0.0654,0.0510)&-0.0500 & 0.1188 & (-0.3161,0.1643)& -0.1021 & 0.1607 & (-0.4618,0.1463)\\
&   $\beta_2$ & -0.0035 & 0.0320 & (-0.0806,0.0518)  &  -0.0024 & 0.0269 & (-0.0523,0.0453)& -0.0315 & 0.0797 & (-0.2119,0.1106)& -0.0296 & 0.0869 & (-0.2227,0.1394)\\
&   $\beta_3$ & -0.0013 & 0.0401 & (-0.0815,0.0721) & -0.0000 & 0.0328 & (-0.0469,0.0612)&-0.0189 & 0.1189 & (-0.2774,0.2294)&  -0.0138 & 0.1347 & (-0.3024,0.2697)\\
&   $\lambda_1$ & 0.7527 & 0.1127 & (0.5384,0.9866) & 0.7527 & 0.1109 & (0.5449,0.9824)& 0.8084 & 0.1199 & (0.5786,1.0431)&  0.8229 & 0.1388 & (0.5556,1.0893)\\
&   $\lambda_2$ &  0.9129 & 0.1439 & (0.6299,1.2102)  & 0.9137 & 0.1371 & (0.6586,1.2114)&0.9833 & 0.1416 & (0.7151,1.2678)&  1.1475 & 0.2447 & (0.7205,1.6404)\\
&   $\lambda_3$ &  0.6876 & 0.1408 & (0.4214,0.9816) & 0.6875 & 0.1384 & (0.4224,0.9779)& 0.7615 & 0.1343 & (0.4849,1.0198)&  1.0590 & 0.3720 & (0.5449,1.8070)\\   
&   $\lambda_4$ &  0.5760 & 0.1756 & (0.2370,0.9579) &  0.5783 & 0.1749 & (0.2397,0.9474)&0.6583 & 0.1538 & (0.3617,0.9589)&  1.1444 & 0.6581 & (0.3252,2.4241)\\   
&   $\pi(\bm z)$ & 0.6736 & 0.0605 & (0.5496,0.7450) & 0.6546 & 0.0588 & (0.5568,0.7503)&0.6273 & 0.0343 & (0.5618,0.6952)& 0.6666 & 0.0583 & (0.5609,0.7815)\\ 
\bottomrule
\end{tabular} }
\label{tab:Bayes_estimates_E1690data}
\end{sidewaystable}

\clearpage
\subsection{Colon cancer clinical trial data}

As a second dataset for showcasing the applicability of our proposed models, we use clinical trial data on colon cancer reported by the North Central Cancer Treatment Group (\cite{moertel1990levamisole}).
This dataset, publicly available as \textit{colon} in the \texttt{survival} R package \cite{survival-Rpackage}, has previously been analyzed by \cite{barriga2019new} and \cite{cancho2021bayesian} to illustrate promotion time cure and bounded cumulative hazard cure rate models, respectively.

The trial evaluated the efficacy of Fluorouracil (5-FU) combined with Levamisole, as well as Levamisole alone, in preventing recurrence among patients with stage C colorectal cancer following complete surgical resection. A total of $929$ patients were enrolled and followed for a median of seven years. 
After excluding individuals with incomplete information or missing observation times, $n=888$ patients remained. 
For comparability, we use the same dataset and covariates as in \cite{barriga2019new} and \cite{cancho2021bayesian}.
However, when examining time to disease recurrence -- the response variable of interest -- we identified seven patients with recurrence times under $30$ days. These were excluded to remove potentially uninformative early events, leaving $n=881$ patients for the final analysis (with approximately $50\%$ censoring). The recurrence times were converted from days to years by dividing by $365.25$.

\begin{figure}[!htbp]
\centering
\includegraphics[width=1\textwidth]{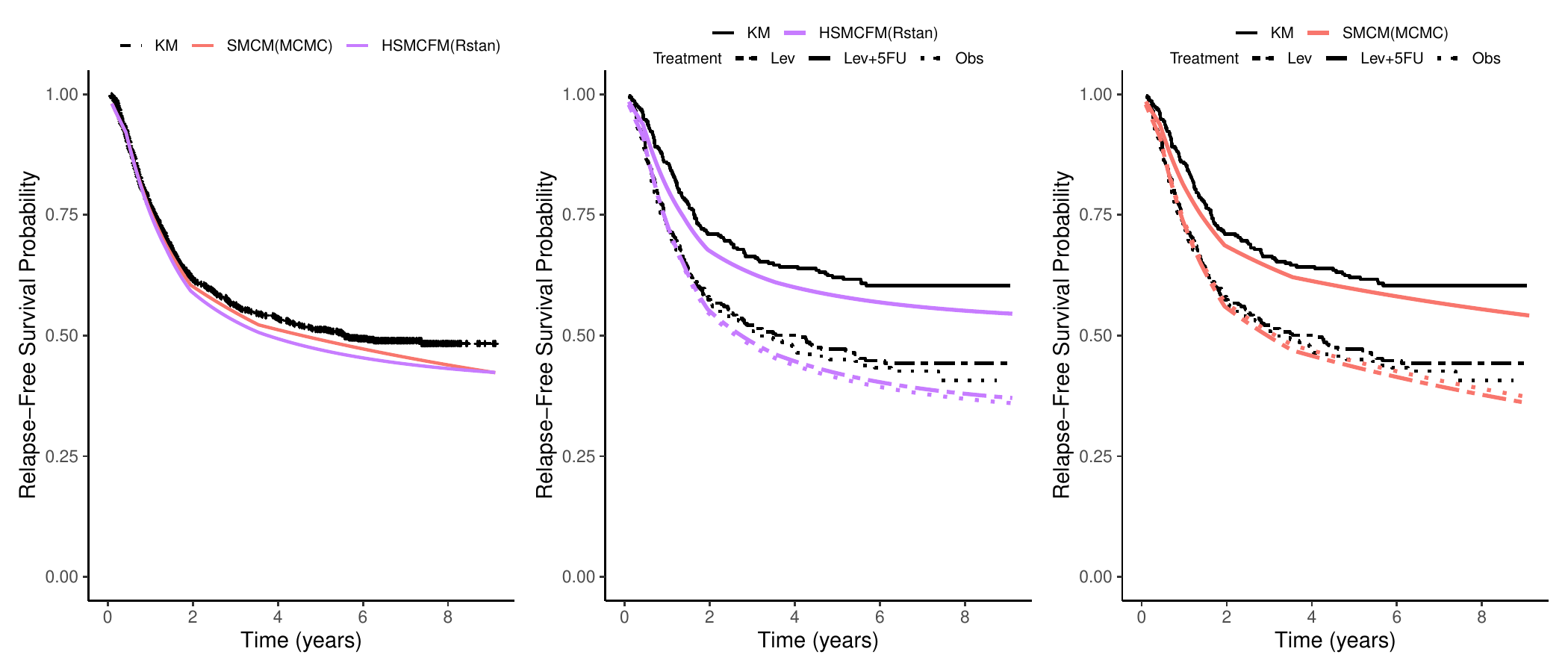}
\caption{Results of the analysis of the colon dataset.
Kaplan–Meier estimates and posterior survival functions obtained from the selected best model HSMCFM(RStan) ($J=7$), and an alternative SMCM(MCMC) ($J=7$) model. 
The left panel shows the overall survival curves, while the middle and right panels display survival curves stratified by treatment, based on the HSMCFM(RStan) and SMCM(MCMC) models, respectively. 
Treatment levels are Observation (Obs) (dotted line), Levamisole (Lev) (dashed line), and  Levamisole+5-FU (Lev+5FU) (solid line).
}
\label{Data2_KM_Model_plots}
\end{figure}

The variables considered in this application are as follows. The outcome of interest is the observed time to disease recurrence, $t$, measured in years (mean $3.9$), together with the censoring indicator ($\delta$), which corresponds to $440$ right-censored cases. Treatment assignment ($z_{1}$) distinguishes three groups: $304$ patients under observation, $291$ receiving Levamisole, and $286$ receiving Levamisole+5-FU. The extent of local spread ($z_{2}$) was recorded at four levels: submucosa ($19$ patients), muscle ($101$), serosa ($724$), and contiguous structures ($37$). The time from surgery to registration ($z_{3}$) was categorized as short ($646$ patients) or long ($235$), and the number of positive lymph nodes ($z_{4}$) was coded as no more than four ($648$) versus more than four ($233$). We use two common covariates, $x_{1}=z_{3}$, $x_{2}=z_{4}$ for the latency part. In addition, we include age ($x_{3}$), dichotomized as under $60$ years ($389$) versus $60$ years or older ($492$), and sex ($x_{4}$), with $426$ females and $455$ males. 
Categorical covariates with more than two levels are coded as dummy variables in the analysis.
Among patients without recurrence, both the mean and median follow-up exceeded six years. 
Figure \ref{Data2_KM_Model_plots} shows the KM survival estimates, both overall and stratified by treatment group -- Observation (Obs), Levamisole (Lev), and Levamisole + 5-FU (Lev+5FU).
A clear plateau appearing around six years suggests the presence of long-term survivors, providing strong motivation for modeling this dataset within a cure rate framework.

\begin{figure}[!h]
\centering
\includegraphics[width=1\textwidth]{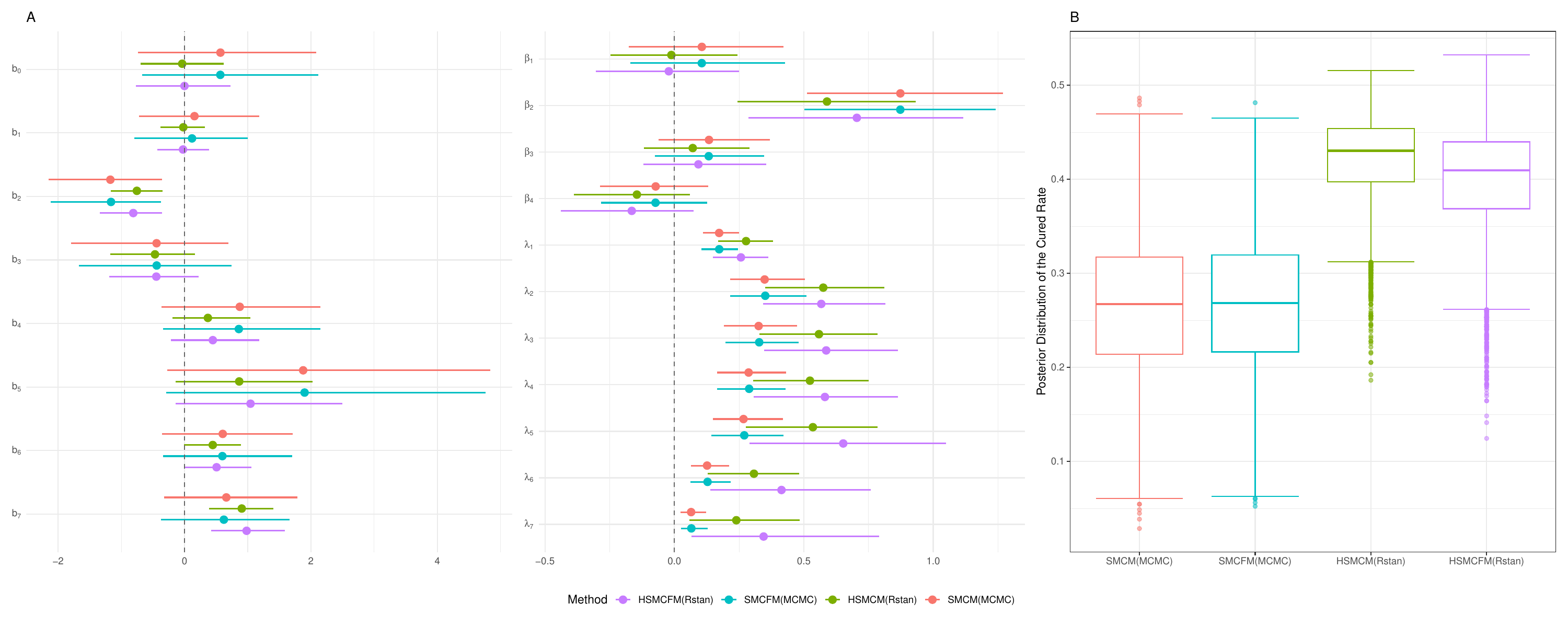}
\caption{Results of the analysis of the colon dataset. Bayesian posterior mean estimates of the parameters for the selected model HSMCFM(RStan) ($J=7$). 
(A) Bayesian posterior mean estimate along with the $95 \%$ HPD credible interval for the parameters $(\bm b, \bm \beta, \bm \lambda)$.
(B) Boxplots of the
posterior means of the cure rate, $1-\pi(\bm z)$, for all patients.}
\label{Data2_selected_model_combined_plots}
\end{figure}

We applied four proposed mixture cure model approaches -- SMCM (MCMC), SMCFM (MCMC), HSMCM (RStan), and HSMCFM (RStan) -- to this dataset.
We consider $J \in {\{3,5, 7, 10\}}$ in the semiparametric structure and run five MCMC chains with $60000$ iterations each, including $10000$ burn-in iterations and applying thinning by $50$. This yields a total of $5000$ posterior samples for each method.
Model comparisons indexes based on BIC, DIC, and LPML obtained from our models, together with the proposed best parametric cure model with frailty from \cite{cancho2021bayesian}, are summarized in Table \ref{tab:Bayes_comparisons_colon_data}. 
While BIC tends to strongly penalize model complexity and may therefore favor overly simplistic specifications in semiparametric mixture cure models, DIC and LPML are more suitable in this context. Both are computed directly from posterior MCMC samples and thus capture the full Bayesian inference.
Hence, we primarily rely on DIC and LPML when selecting the best-fitting model.
From Table \ref{tab:Bayes_comparisons_colon_data}, according to the DIC and LPML criteria, the best-fitting models correspond to the case $J=7$ across all approaches, including the \cite{cancho2021bayesian} model, with the HSMCFM(RStan) showing the best overall performance. 
Figure \ref{Data2_KM_Model_plots} also shows the survival curves of the best model for the case 
$J=7$ -- the HSMCFM(RStan) model -- together with the alternative SMCM(MCMC) model, based on posterior survival estimates for both overall survival and survival stratified by treatment.
Both models provide a good fit to the observed data, closely following the KM estimates during the early follow-up period.
The stratified plots indicate that treatment effects are well captured, with Lev+5FU showing slightly higher survival probabilities than Lev or Observation, which is consistent with the findings of \cite{barriga2019new} and \cite{cancho2021bayesian}.
Overall, the close agreement with the KM estimates and the clear plateau pattern suggest that both models, and especially the HSMCFM(RStan), provide an adequate and realistic fit to the colon dataset.
Posterior summaries for these models are displayed in Figure \ref{Data2_selected_model_combined_plots}, which includes error-bar plots of the posterior mean estimates with $95\%$ credible intervals for the regression coefficients and boxplots of the posterior cure rate estimates.

\begin{table}[!h]
\centering
 \renewcommand{\arraystretch}{1.30}
 \caption{Results of the analysis of the colon dataset. Bayesian model comparison criteria for all fitted models.}
  \resizebox{1 \textwidth}{!}{
\begin{tabular}{ccccccc}
\toprule 
$J$ & Criteria & SMCM(MCMC) & SMCFM(MCMC) & HSMCM(RStan) & HSMCFM(RStan) \\ \hline 
3 & BIC &  \textbf{2433.492} & 2440.787&  2434.239& 2442.546\\
  & DIC &  2345.555& 2346.61& 2345.856& 2348.239\\ 
  & LPML&  -1172.814& -1173.318& -1172.904& -1173.601\\ \hline
5 & BIC & 2437.192 & 2443.363 & 2434.122 & 2440.867 \\
  & DIC & 2334.055 & 2334.022 & 2333.185 & 2330.522 \\
  & LPML& -1168.952 & -1168.487& -1166.915 & -1166.847\\ \hline  
7 & BIC &  2443.446 &2449.939& 2439.021 &  2445.281\\ 
  & DIC &  2328.424  & 2328.535 & 2326.446& \textbf{2321.252}\\ 
  & LPML&  -1165.443 & -1166.153 & -1163.592  & \textbf{-1163.40} \\\hline   
10& BIC & 2472.541 &2479.028 & 2464.70 & 2470.557 \\
  & DIC & 2338.989 & 2340.28& 2334.747 & 2326.428\\ 
  & LPML& -1172.458 & -1172.158& -1167.704 & -1167.411\\\hline  
\multicolumn{3}{c}{\shortstack{Parametric BCH-PVF \\ model in \cite{cancho2021bayesian} } } & DIC: 2328.79 & LPML: -1164.865 & \\ \hline
\end{tabular}
}
\label{tab:Bayes_comparisons_colon_data}
\end{table}

\newpage
\section{Discussion and Conclusion}\label{sec:conclusion}

In this study, we developed a hierarchical Bayesian framework for semiparametric mixture cure models, which allows flexible inclusion of a frailty component to account for unobserved heterogeneity among patients. The proposed approach integrates a hierarchical structure inspired by the Bayesian Lasso, enabling shrinkage and regularization of regression coefficients while preserving interpretability. Through extensive simulation studies across a range of scenarios, we demonstrated that the models provide accurate and robust estimates of both the cured fraction and the survival distribution among uncured individuals.

Our simulation results can be summarized from two complementary perspectives: model structure (whether a frailty term is included) and algorithmic implementation (our MCMC versus \texttt{RStan}). When the number of observations and the number of intervals for the semiparametric hazard function are sufficient, both model structures -- SMCM and SMCFM -- showed comparable performance, regardless of whether the data were generated under SMCM or parametric MCM settings.
Bayesian model selection criteria further confirmed the advantages of the proposed methods, particularly under model misspecification scenarios in both simulation and real-data analyses. The models implemented using our MCMC algorithm, namely SMCM(MCMC) and SMCFM(MCMC), performed as well as the \texttt{RStan}-based HSMCM(RStan) model. Although the HSMCFM(RStan) model generally showed slightly less favorable performance compared to other models, it still produced comparable estimates, particularly for the uncured rate.

The application to two well-known datasets -- the E1690 melanoma trial and the colon cancer trial -- illustrated the practical utility of the proposed framework. In both datasets, the models successfully captured long-term survival dynamics and identified cured subgroups with reasonable precision. In the colon cancer data, some differences were observed between our MCMC-based and \texttt{RStan}-based implementations; such discrepancies can reasonably be attributed to the greater heterogeneity and censoring patterns present in this real dataset, rather than to differences in model performance. 
Importantly, the flexibility to include or exclude the frailty component allows the model to adapt to the underlying data structure, providing more reliable inference in complex clinical settings. These results highlight the models’ capacity to generate meaningful insights from real-world survival data, even in the presence of heterogeneity and censoring.

Despite these advantages, some limitations should be acknowledged. The computational cost of our MCMC method is higher than that of models implemented in \texttt{RStan} in most cases. We plan to reduce this cost by integrating C++ implementations within our MCMC code, via the use of the \texttt{Rcpp} R package \cite{eddelbuettel2011rcpp}. Although \texttt{RStan} provides a highly efficient and automated inference framework for models of moderate complexity, its scalability becomes limited as the number of covariates increases. In high-dimensional settings, such as those involving hundreds of predictors or penalized regression structures, \texttt{RStan} can face convergence difficulties, long sampling times, and high memory demands.
Indeed, the use of shrinkage priors promotes parsimony by controlling overfitting, which is particularly advantageous when dealing with a large number of covariates or high-dimensional data.
Our proposed MCMC algorithm, differently from the \texttt{RStan} implementation, is designed to handle hierarchical and shrinkage priors effectively, demonstrating promising stability and computational feasibility even in more complex or high-dimensional configurations. Similar MCMC-based approaches have already been successfully proposed in high-dimensional survival analysis contexts, for example \cite{lee2011bayesian, lee2015survival} and the R package by \cite{psbcSpeedUp-Rpackage}. This similar line of research suggests that the framework proposed in the present paper can be naturally extended to high-dimensional survival data, including omics or other information-rich covariate datasets. Detailed investigations in this direction will be the focus of our future research.

One of the important outcomes of this study is the development of an R package -- currently available on our GitHub repository -- that implements all proposed models. We plan to extend this package with additional functionalities in future work, and the forthcoming version will be submitted to CRAN to ensure wider accessibility and facilitate its use in research and clinical practice.

In conclusion, the hierarchical Bayesian semiparametric mixture cure model introduced here provides a flexible, robust, and interpretable framework for survival analysis in the presence of cured patients. 
By integrating hierarchical shrinkage and frailty components within a unified Bayesian structure, the method captures both observed and unobserved sources of variation, leading to improved inference and predictive performance.
Therefore, this framework contributes to the growing toolkit of Bayesian survival models and offers practical utility for researchers and clinicians seeking to better understand long-term survival and cure probabilities in medical studies.

\section*{Acknowledgements}
Computer simulations were performed in HPC solutions provided by the Oslo Centre for Biostatistics and Epidemiology, at the Department of Biostatistics, Institute of Basic Medical Sciences, University of Oslo.

\section*{Funding}
This project has received funding from the European Union’s Horizon 2020 research and innovation programme under the Marie Skłodowska-Curie grant agreement No 801133.

\section*{Conflict of interest statement}
None declared.

\section*{Software and reproducibility}
An R package implementing the proposed model algorithms and enabling replication of the simulation studies and real data analyses can be installed directly from our GitHub repository via \texttt{devtools::install\_github( \\ "fatihki/BayesSMCM")}.

\section*{Appendix}

\begin{algorithm}[H]
\caption{Metropolis-within-Gibbs algorithm for SMCFM}
\label{alg:MCMC_SMCFM}
\begin{algorithmic}[1]
\State Initialize $\bm{b}^{(0)}, \; \bm{\beta}^{(0)},\; \bm \lambda^{(0)}, \; \theta^{ (0)}, \; \bm{\tau}^{(0)}, \bm{\tau}^{*(0)},\sigma^{2(0)}, \; \sigma^{*2(0)},\;  \eta^{2(0)}$ and $ \eta^{*2(0)}$
\For{$g = 1,\dots, M$} 
  \For{$k = 1, \dots, p_1+1$} \Comment{Update $\bm{b}$ by Metropolis–Hastings algorithm based on (\ref{posterior_bk_SMCFM})}
    \State Sample $b_k^{(prop)} \sim \mathcal{N}(b_k^{(prop)} | \mu_{b_k}^{(g-1)}, 1 )$ and $u_k \sim \mathcal{U}(0,1)$.
    \State Calculate the acceptance probability: 
        \[
            p_k = \frac{\pi( b_{k}^{(prop)} \mid \mathbf{b}^{(-k)(g-1)}, \bm{\beta}^{(g-1)},  \bm{\lambda}^{(g-1)}, \theta^{(g-1)}, \bm{\tau}^{(g-1)}, \sigma^{2 (g-1)}, \eta^{2 (g-1)} , \mathbf{D}) \; / \; J_{g}(b_k^{(prop)} \mid b_k^{(g-1)})} 
            {\pi( b_{k}^{(g-1)} \mid \mathbf{b}^{(-k)(g-1)}, \bm{\beta}^{(g-1)},  \bm{\lambda}^{(g-1)}, \theta^{(g-1)}, \bm{\tau}^{(g-1)}, \sigma^{2 (g-1)}, \eta^{2 (g-1)} , \mathbf{D}) \; / \; J_{g}(b_k^{(g-1)} \mid b_k^{(prop)})} 
        \]
        \State \textbf{if} $u_k < p_k$ \textbf{then} $b_k^{(g)} = b_k^{(prop)}$ \textbf{else} $b_k^{(g)} = b_k^{(g-1)}$
  \EndFor 
   \For{$k = 1, \dots, p_2$} \Comment{Update $\bm{\beta}$ by Metropolis-Hastings algorithm based on  (\ref{posterior_betak_SMCFM})}
      \State Sample proposal $\beta_k^{(prop)} $ from $\mathcal{N}(\beta_k^{(prop)} \mid \mu_{\beta_k}^{(g-1)}, 1 )$ and $u_k \sim \mathcal{U}(0,1)$.
        \State Calculate the acceptance probability: 
        \[
            p_k = \frac{\pi( \beta_{k}^{(prop)} \mid \bm{\beta}^{(-k)(g-1)}, \bm{b}^{(g-1)},  \bm{\lambda}^{(g-1)}, \theta^{(g-1)}, \bm{\tau}^{*(g-1)}, \sigma^{*2 (g-1)}, \eta^{*2 (g-1)} , \mathbf{D}) \; / \; J_{g}(\beta_k^{(prop)} \mid \beta_k^{(g-1)})} 
            {\pi( \beta_{k}^{(g-1)} \mid \bm{\beta}^{(-k)(g-1)}, \bm{b}^{(g-1)},  \bm{\lambda}^{(g-1)}, \theta^{(g-1)}, \bm{\tau}^{*(g-1)}, \sigma^{*2 (g-1)}, \eta^{*2 (g-1)} , \mathbf{D}) \; / \; J_{g}(\beta_k^{(g-1)} \mid \beta_k^{(prop)})} 
        \]
    \State \textbf{if} $u_k < p_k$ \textbf{then} $\beta_k^{(g)} = \beta_k^{(prop)}$ \textbf{else} $\beta_k^{(g)} = \beta_k^{(g-1)}$
   \EndFor 
      \For{$k = 1, \dots, J$} \Comment{Update $\bm{\lambda}$ by Metropolis-Hastings algorithm based on (\ref{posterior_lambdak_SMCFM})}
      \State Sample proposal $\lambda_k^{(prop)}$ from $\text{Gamma}(\lambda_k^{(prop)} \mid a_{\lambda_k}^{(g-1)}, 1 )$ and $u_k \sim \mathcal{U}(0,1)$.
      \State Calculate the acceptance probability: 
        \[
          p_k = \frac{\pi( \lambda_{k}^{(prop)} \mid \bm{\lambda}^{(-k)(g-1)}, \bm{b}^{(g-1)}, \bm{\beta}^{(g-1)}, \theta^{(g-1)}, \mathbf{D})  / J_{g}(\lambda_k^{(prop)} \mid \lambda_k^{(g-1)}) } 
          {\pi( \lambda_{k}^{(g-1)} \mid \bm{\lambda}^{(-k)(g-1)}, \bm{b}^{(g-1)}, \bm{\beta}^{(g-1)}, \theta^{(g-1)}, \mathbf{D}) / J_{g}(\lambda_k^{(g-1)} \mid \lambda_k^{(prop)}) } 
       \]
      \State \textbf{if} $u_k < p_k$ \textbf{then} $\lambda_k^{(g)} = \lambda_k^{(prop)}$ \textbf{else} $\lambda_k^{(g)} = \lambda_k^{(g-1)}$
   \EndFor 
  \Statex \Comment{Update $\theta$ by Metropolis-Hastings algorithm based on (\ref{posterior_theta_SMCFM})}
  \State Sample proposal $\theta^{(prop)}$ from $\text{Gamma}(\theta^{(prop)} \mid a_{\theta}^{(g-1)}, 1 )$ and $u \sim \mathcal{U}(0,1)$. 
      \State Calculate the acceptance probability: 
        \[
          p = \frac{\pi( \theta^{(prop)} \mid \bm{b}^{(g-1)}, \bm{\beta}^{(g-1)}, \bm{\lambda}^{(g-1)}, \mathbf{D})  / J_{g}(\theta^{(prop)} \mid \theta^{(g-1)}) } 
          {\pi( \theta^{(g-1)} \mid \bm{b}^{(g-1)}, \bm{\beta}^{(g-1)}, \bm{\lambda}^{(g-1)},  \mathbf{D}) / J_{g}(\theta^{(g-1)} \mid \theta^{(prop)}) } 
       \]
      \State \textbf{if} $u < p$ \textbf{then} $\theta^{(g)} = \theta^{(prop)}$ \textbf{else} $\theta^{(g)} = \theta^{(g-1)}$
  \State Sample $1/\tau_{k}^2, \; k=1,\cdots, p_1 $ from (\ref{posterior_tau}).
  \State Sample $1/\tau_{k}^{*2}, \; k=1, \cdots, p_2$ from  (\ref{posterior_tau_star}).
  \State Sample $\sigma^2 $ from (\ref{posterior_sigma}).
  \State Sample $\sigma^{*2} $ from (\ref{posterior_sigma_star}).
  \State Sample $\eta^2 $ from (\ref{posterior_eta}).
  \State Sample $\eta^{*2} $ from (\ref{posterior_eta_star}).
\EndFor 
\end{algorithmic}
\end{algorithm}

\begin{algorithm}[H]
\caption{HSMCFM(RStan) algorithm}
\label{alg:HSMCFM_Rstan}
\begin{algorithmic}[1]
\State Specify the likelihood based on (\ref{lik_SMCFM}) and hierarchical priors such as: 
\[ \begin{aligned}
& \lambda_j \sim \text{Gamma}(a,b), \;\; j=1,\cdots,J, \;\; \theta \sim \text{Gamma}(\theta_a,\theta_b), \\
& b_j \mid \tau_{j}^2, \sigma^2 \sim \mathcal{N}(0, \sigma^2 \tau_{j}^2), \;\;
 \tau_{j}^2 \mid \eta^2 \sim \text{Exponential}(\eta^2/2), \;\; j=1,\cdots,p_1+1, \\
 & \eta^2 \sim \text{Gamma}( r_1, \delta_{1}), \;\; \sigma^2  \sim \mathcal{U}(0, 1000), \\
&\beta_j \mid \tau_{j}^{*{2}}, \sigma^{*2} \sim \mathcal{N}(0, \sigma^{*2} \tau_j^{*{2}}), \;\;\ 
 \tau_{j}^{*{2}} \mid \eta^{*{2}} \sim \text{Exponential}(\eta^{*{2}}/2), \;\; j=1,\cdots,p_2, \\
 & \eta^{*{2}} \sim \text{Gamma}( r_2, \delta_{2}), \;\; \sigma^{*2}  \sim \mathcal{U}(0, 1000).  \
\end{aligned}
\]
\State Defining the joint posterior $\pi(\bm{b}, \bm{\beta}, \bm{\lambda}, \theta, \bm{\tau}, \bm{\tau}^*, 
\sigma^2, \sigma^{*2}, \eta^2, \eta^{*2} \mid \bm{D})$.
\State Implement the model in \texttt{Stan} and provide data $\bm{D}$.
\State Compile the model in \texttt{RStan}.
\State Run the No-U-Turn Sampler (NUTS) for $M$ iterations.
\State Collect posterior draws of $(\bm b, \bm \beta, \bm \lambda,\theta)$, and compute summaries (posterior means, credible intervals, and diagnostics).
\end{algorithmic}
\end{algorithm}

\begin{algorithm}[H]
\caption{SMCFM(RStan) algorithm}
\label{alg:SMCFM_Rstan}
\begin{algorithmic}[1]
\State Specify the likelihood based on (\ref{lik_SMCFM}) and priors such as: $\mathbf{b} \sim N(\mathbf{0}, \bm{\Sigma_{b}})$, $\bm{\beta} \sim N(\mathbf{0}, \bm{\Sigma_{\beta}})$, $\bm\lambda \sim Gamma(a,b)$, $\theta \sim Gamma(\theta_a,\theta_b)$.
\State Defining the joint posterior $\pi(\bm{b}, \bm{\beta}, \bm{\lambda}, \theta, \bm{\tau}, \bm{\tau}^*, 
\sigma^2, \sigma^{*2}, \eta^2, \eta^{*2} \mid \bm{D})$.
\State Implement the model in \texttt{Stan} and provide data $\bm{D}$.
\State Compile the model in \texttt{RStan}.
\State Run the NUTS for $M$ iterations.
\State Collect posterior draws of $(\bm b, \bm \beta, \bm \lambda,\theta)$, and compute summaries (posterior means, credible intervals, and diagnostics).
\end{algorithmic}
\end{algorithm}

\bibliographystyle{unsrtnat}

\end{document}